\documentclass[twocolumn]{aastex63}
\usepackage[utf8]{inputenc}
\usepackage{booktabs}
\usepackage{amsmath}
\usepackage{float}
\usepackage{fontawesome5}

\newcommand{\new}[1]{{\color{teal} #1}}
\DeclareUnicodeCharacter{FF0C}{é}
\newcommand{\name}{Photo-$z$SNthesis}
\usepackage{hyperref}

\makeatletter
\newcommand{\github}[1]{%
   \href{#1}{\faGithub}%
}
\makeatother

\begin{document}
\title{Photo-$z$SNthesis: Converting Type Ia Supernova Lightcurves to Redshift Estimates via Deep Learning}

\author[0000-0003-1899-9791]{Helen Qu}
\affiliation{Department of Physics and Astronomy, University of Pennsylvania, Philadelphia, PA 19104, USA}
\author[0000-0003-2764-7093]{Masao Sako}
\affiliation{Department of Physics and Astronomy, University of Pennsylvania, Philadelphia, PA 19104, USA}

\correspondingauthor{Helen Qu}
\email{helenqu@sas.upenn.edu}

\begin{abstract}
Upcoming photometric surveys will discover tens of thousands of Type Ia supernovae (SNe Ia), vastly outpacing the capacity of our spectroscopic resources. In order to maximize the science return of these observations in the absence of spectroscopic information, we must accurately extract key parameters, such as SN redshifts, with photometric information alone. We present \name, a convolutional neural network-based method for predicting full redshift probability distributions from multi-band supernova lightcurves, tested on both simulated Sloan Digital Sky Survey (SDSS) and Vera C. Rubin Legacy Survey of Space and Time (LSST) data as well as observed SDSS SNe. We show major improvements over predictions from existing methods on both simulations and real observations as well as minimal redshift-dependent bias, which is a challenge due to selection effects, e.g. Malmquist bias. Specifically, we show a $61\times$ improvement in prediction bias $\langle \Delta z \rangle$ on PLAsTiCC simulations and $5\times$ improvement on real SDSS data compared to results from a widely used photometric redshift estimator, LCFIT+Z. The PDFs produced by this method are well-constrained and will maximize the cosmological constraining power of photometric SNe Ia samples. 
\end{abstract}

\section{Introduction}
The study of Type Ia supernovae (SNe Ia) has proven to be a crucial tool in modern cosmology, providing insight into the expansion rate of the universe and the properties of dark energy \citep{riess, perlmutter}. Measuring the cosmological redshift of each SN, a proxy quantity for recessional velocity, is essential for accurate estimation of the distance-redshift relation and resulting cosmological analyses. However, traditional methods of measuring redshifts are time-consuming and resource-intensive, primarily relying on spectroscopic observations of the SNe themselves or their host galaxies. Using host galaxy redshifts can also lead to cosmological biases if the host is incorrectly identified (Qu et al., in prep).
\begin{figure*}
    \centering
    \includegraphics[scale=0.45]{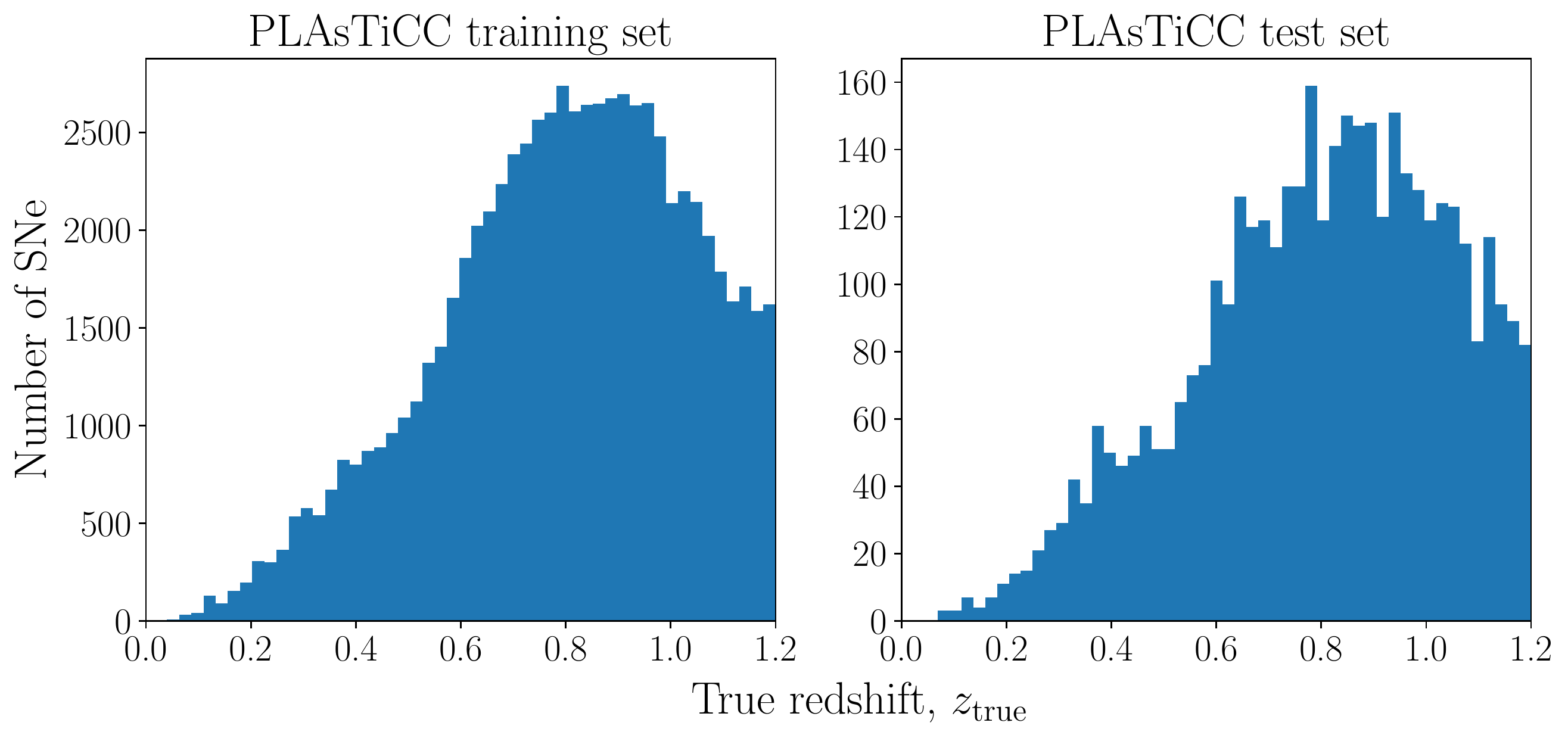}
    \caption{True redshift distribution of the 73,620 simulated SNe Ia in the LSST (PLAsTiCC) training dataset (left) and 4,057 simulated SNe Ia in the test dataset (right).}
    \label{fig:plasticc-z-dist}
\end{figure*}

Most SNe Ia cosmological analyses so far have relied on SN spectra for SN type confirmation as well as redshift information, but only with samples of up to $\sim 1,500$ SNe \citep[e.g.][]{Abbott_2019, pantheonplus}. Spectroscopic follow-up of all SNe Ia candidates or their host galaxies will become infeasible with future sky surveys such as the Legacy Survey of Space and Time at the Vera C. Rubin Observatory (LSST), which will discover tens of thousands of SNe Ia over the course of their observational lifetimes \citep{lsst_book}. Recent improvements in photometric SN classification \citep[e.g.][]{supernnova, scone} have drastically reduced the likelihood of non-Ia contamination in photometric SNe Ia samples and enabled cosmological analyses with photometrically classified samples \citep{vincenzi_contamination}. However, spectroscopic redshifts were still available for the host galaxies of these photometrically confirmed SNe Ia and were used as the SN redshifts. SN redshift estimates independent of host galaxy redshift can also serve as an independent cross-check for host galaxy association, ensuring accurate studies of host galaxy correlations and corrections for the mass step \citep[e.g.][]{Rigault_2020}. Accurate photometric redshift estimates for SNe that are independent of host galaxy spectroscopic redshifts are thus the final building block required to enable SN Ia cosmology for the LSST era.



Cosmological inference frameworks that account for the inflated uncertainties from photometric redshifts are currently being developed. \citet{mitra2022, Dai_2018} show promising results with simulated LSST samples and SN photometric redshifts fitted using host galaxy redshift priors. In particular, \citet{Dai_2018} recovers a fitted $\Omega_m$ value consistent with the input cosmology when using SN photo-$z$s fitted with a host galaxy photo-$z$ prior. \citet{mitra2022} shows a 2\% effect on fitted cosmological parameters of an assumed systematic uncertainty due to the use of SN photo-$z$s of 0.01. Using observed data from the Dark Energy Survey, \citet{chen2022} performed a cosmological analysis using a subset of $\sim 100$ SNe Ia hosted by galaxies in the redMaGiC catalog, which have both photometric and spectroscopic redshifts. The difference in best-fit cosmological parameters between using spectroscopic and photometric redshifts was found to be minimal, $\Delta w \sim 0.005$. redMaGiC galaxies were chosen for this analysis due to their particularly well-constrained photometric redshift estimates, which is not representative of the full population of SN host galaxies. However, even with a sound cosmological inference framework, galaxy photometric redshifts are often inaccurate or plagued with large uncertainties, and requiring host galaxy information to produce SN photometric redshift estimates may introduce additional biases. 

These issues with traditional redshift determination along with the development of a cosmological framework for photometric redshifts has led to a growing interest in alternative techniques for predicting redshifts for Type Ia supernovae. In this work, we intoduce a novel machine learning algorithm to predict full redshift probability distributions for Type Ia supernovae based solely on lightcurve data. Our estimator harnesses the constraining power on redshift of the SN lightcurves and can additionally provide an independent cross-check on host galaxy matches, minimizing mismatch rates. We present a detailed analysis of our model, including its accuracy and limitations, and discuss the potential implications of our findings for future cosmological studies.

\subsection{Photo-z Estimation}

Most of the existing literature on photometric redshift estimation is on galaxy photo-$z$s. These approaches generally use either (1) a training set of galaxy photometric observables, such as colors and magnitudes, to determine a mapping to spectroscopic redshifts \citep[e.g.][]{Brunner_1997}; or (2) template fitting, in which observed properties are compared with redshifted template spectra to determine the best fit redshift value \citep[e.g.,][]{Benitez_2000}. There have also been successful machine learning-based galaxy photo-$z$ models developed, e.g. \citet{disanto, Pasquet_2018} using convolutional neural networks on galaxy images, and \citet{SOM} using a self-organizing map to relate color-magnitude space and redshifts.

\begin{figure*}
    \centering
    \includegraphics[scale=0.45]{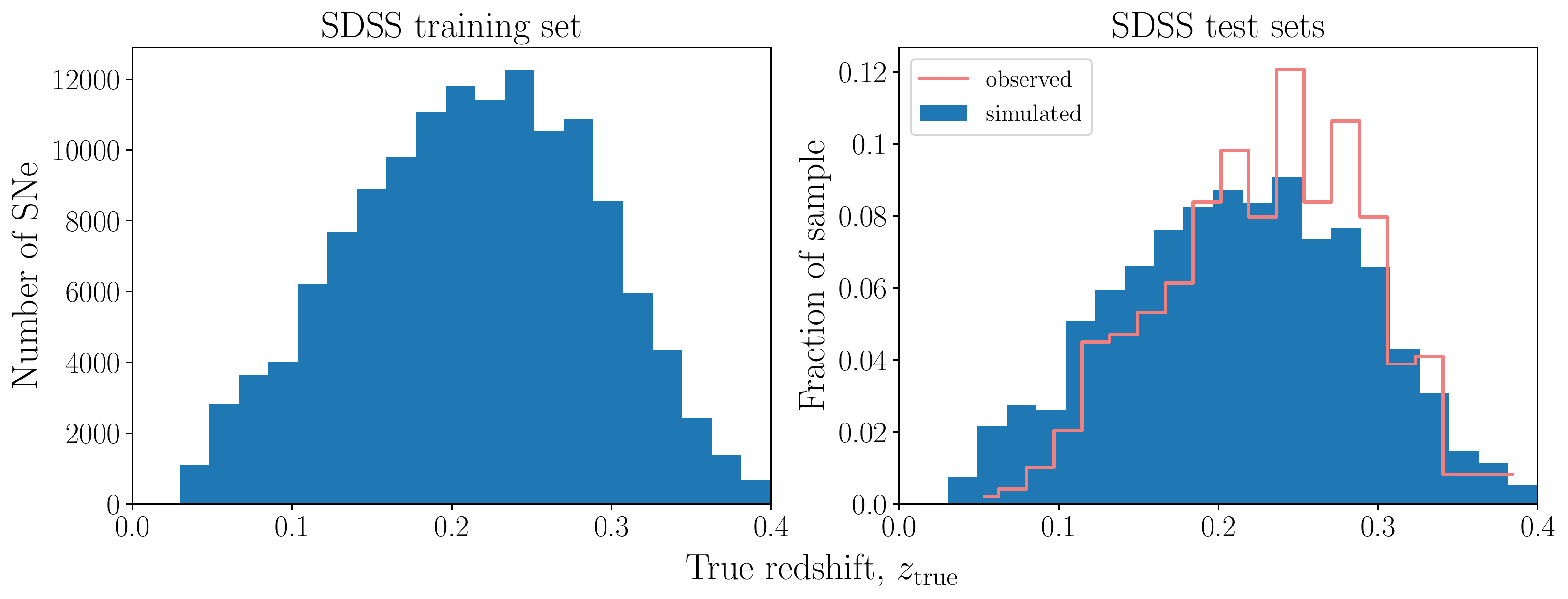}
    \caption{True redshift distribution of the 135,521 simulated SNe Ia in the SDSS training dataset (left) and the 5,274 simulated SNe Ia and 489 observed photometrically confirmed SNe Ia in the SDSS test sets (right).}
    \label{fig:sdss-z-dist}
\end{figure*}
Photometric redshift estimation for SNe uses many of the same techniques. \citet{Kim_2007} used the SALT2 lightcurve model \citep{salt2} to determine photometric redshifts and distances as well as uncertainties using the Fisher information matrix. \citet{Palanque_Delabrouille_2010} and \citet[][LCFIT+Z]{Kessler_2010} extended the SALT2 lightcurve fit to include redshift as a fitted parameter, and incorporating the host galaxy redshift as a prior. These approaches find the best fit redshift by comparing photometric observables to those expected from the SALT2 model via $\chi^2$ minimization. An analytic approach was presented in \citet{wang2015} which assumed a functional form for the redshifts as a function of multi-band photometric fluxes and fit for free parameters using a training set of SNe Ia with spectroscopic redshifts. \citet{deOliveira:2022tvw} applied machine learning techniques to this problem, performing regression using features obtained from principal component analysis. However, many of these results suffer from redshift-dependent bias, in which photo-$z$s for high-redshift SNe are systematically underestimated (see e.g. Figure 6 of \citet{Kessler_2010}, Figure 7 of \citet{deOliveira:2022tvw}). Most of these studies also estimate a redshift value and uncertainty rather than the full probability density function (PDF).

\subsection{Overview}
In this work we present \name, a convolutional neural network model that uses multi-band photometry to predict a full redshift PDF. Our approach uses the raw photometric data and does not require any manual feature engineering to determine the most predictive features. We evaluated \name\ using LSST and SDSS simulated SNe Ia as well as the SDSS photometric SN Ia sample, and show that our redshift predictions have low scatter and suffer from minimal redshift-dependent bias on all tested SN samples. Our results are a promising step towards precision photometric SN Ia cosmology.

We introduce the simulated and observed data samples used for evaluation in \S 2, as well as the preprocessing step used to transform multi-band lightcurves into convolutional neural network inputs. In \S 3, we describe the model architecture and training strategy. We present results, comparisons with LCFIT+Z, and further experiments in \S 4, and conclude in \S 5.

\section{Data}
\subsection{Data Sources}
\label{subsec:data}
We present results on simulated SNe Ia from two surveys: LSST \citep{ivezic_lsst} and the SDSS-II SN survey \citep{sdss}. We also demonstrate that our model generalizes well to an observed photometric SN Ia dataset by showing photo-$z$ predictions on SDSS SNe classified as type Ia by SuperNNova \citep{supernnova}.
We use the SuperNova ANAlysis software \citep[SNANA,][]{snana} to produce all simulated SN lightcurves used in this work.

\subsubsection{Simulated LSST SNe Ia (PLAsTiCC)}
LSST is a ground-based dark energy survey program that will discover millions of SNe over the 10 year survey duration. The 8.4m Simonyi Survey optical telescope at the Rubin Observatory uses a state-of-the-art 3200 megapixel camera with a 9.6 $\text{deg}^2$ field of view that will provide deeper and wider views of the universe with unprecedented quality. LSST will observe nearly half the night sky each week to a depth of $24^{\text{th}}$ magnitude in $ugrizY$ photometric bands spanning wavelengths from ultraviolet to near-infrared.

We simulate LSST-like observations of SNe Ia in \new{$ugrizY$} photometric bands following the model, rates, and LSST observing conditions developed for the PLAsTiCC dataset \citep{plasticc_data, plasticc_models} \new{for $0.05 < z < 1.2$}. We use the SALT2 lightcurve model \citep{salt2} with training parameters derived from the Joint Lightcurve Analysis \citep{jla} extended into the ultraviolet and near-infrared by \citet{hounsell2018} following the procedure described in \citet{pierel2018}. While PLAsTiCC included two LSST observing strategies, the Deep Drilling Fields (DDF) as well as the Wide-Fast-Deep (WFD), we simulate only the DDF subsample. We coadd all observations within the same night, following PLAsTiCC, resulting in observations that are $\sim 2.5$x more frequent and $\sim 1.5$ mag deeper than the WFD sample. Though we focused on the DDF sample for this work, evaluating the performance of \name\ on WFD simulations is an important direction for future studies.

As PLAsTiCC is the name of the dataset we emulated while LSST is the survey we simulate, we will use both interchangeably to denote ``simulated LSST SN lightcurves following the PLAsTiCC dataset".

\subsubsection{Simulated SDSS SNe Ia}
\label{subsec:sdss-sims}
The SDSS-II Supernova Survey identified and measured light curves for intermediate-redshift ($0.05 < z < 0.4$) SNe Ia using repeated five-band ($ugriz$) imaging of Stripe 82, a stripe $2.5^{\circ}$ wide centered on the celestial equator in the Southern Galactic Cap. The primary instrument for this survey is the SDSS CCD camera mounted on a dedicated 2.5m telescope at Apache Point Observatory, New Mexico. Over the three observing seasons between 2005 and 2007, SDSS discovered 10,258 transient and variable objects, with 536 spectroscopically confirmed SNe Ia and an additional 907 photometrically classified SNe Ia candidates \citep{sako2018}. \citet{sdss} provides an in-depth review of the SDSS-II SN survey.

We simulate SDSS SNe Ia in $ugriz$ photometric bands using the SALT3 model described by \citet{Kenworthy_2021} due to its improved wavelength range coverage in the near infrared, allowing $z$ band observations to be simulated for SNe at low redshifts. We additionally extend this model to $500$\AA\ to simulate $u$ band observations at high redshifts. We use simulated observing conditions from \citet{sdss_simlib} and host galaxy spectroscopic detection efficiency from \cite{snana}. We simulate a \textit{photometric} SDSS SNe Ia dataset using the host spectroscopic detection efficiency rather than a SN spectroscopic detection efficiency to demonstrate the performance of \name\ on the practical use case of a photometric sample.

\subsubsection{Observed SDSS SNe Ia}
In addition to simulated SNe, we test \name\ on SDSS lightcurves from \citet{sako2018} classified by SuperNNova \citep{supernnova} as likely SNe Ia, defined as SNe with SNIa probability $P_{\text{Ia}} \geq 0.5$. ``True" redshifts $z_{\text{true}}$ for the sample are from spectra of the SNe themselves or their host galaxies. Details on the training data and training procedure for the SuperNNova model used here can be found in Popovic et al., in prep. 489 lightcurves remain after the selection cuts described in Sections~\ref{subsec:cuts} and~\ref{subsec:preprocess}. The number of SDSS lightcurves remaining after each set of cuts is shown in Table~\ref{tbl:sdss-cuts}. 

\subsubsection{Lightcurve Selection}
\label{subsec:cuts}
We apply basic selection cuts to the simulated LSST sample. We require

\begin{itemize}
    \item at least 5 observations of each SN
    \item signal-to-noise ratio (SNR) $> 3$ for at least one observation each in 2 of the $griz$ filters
\end{itemize}

We apply selection cuts, following \citet{popovic2020}, to both the simulated and observed SDSS data to remove poor quality lightcurves. We define a rest-frame age, $T_{\text{rest}} = (t - t_{\text{peak}})/(1+z)$, where $t$ is the observation date, $t_{\text{peak}}$ is the estimated epoch of SN peak brightness from SNANA, and $z$ is the redshift of the event. We require

\begin{itemize}
    \item $0 < T_{\text{rest}} < 10$ 
    \item SNR $> 5$ for at least one observation each in 2 of the $griz$ filters
\end{itemize}

We additionally require SNe in the SDSS data to have a spectroscopic redshift from either the SN spectrum or the SN host galaxy. 

\begin{table}
    \centering
    \begin{tabular}{l c}
        \toprule
        Cut & Number of SNe \\
        \midrule
        Full sample & 10,258 \\
        Lightcurve selection (\S\ref{subsec:cuts}) & 2,044 \\
        Likely SNe Ia ($P_{\text{Ia}} \geq 0.5$) & 1,037 \\
        SALT fit cuts (\S\ref{subsec:preprocess}) & 555\\
        Successful LCFIT+Z fit & 489 \\
        \bottomrule
    \end{tabular}
    \caption{Number of SDSS lightcurves remaining after each selection cut. We evaluate \name\ on the remaining 489 SNe.}
    \label{tbl:sdss-cuts}
\end{table}

\begin{figure*}
    \centering
    \includegraphics[scale=0.47]{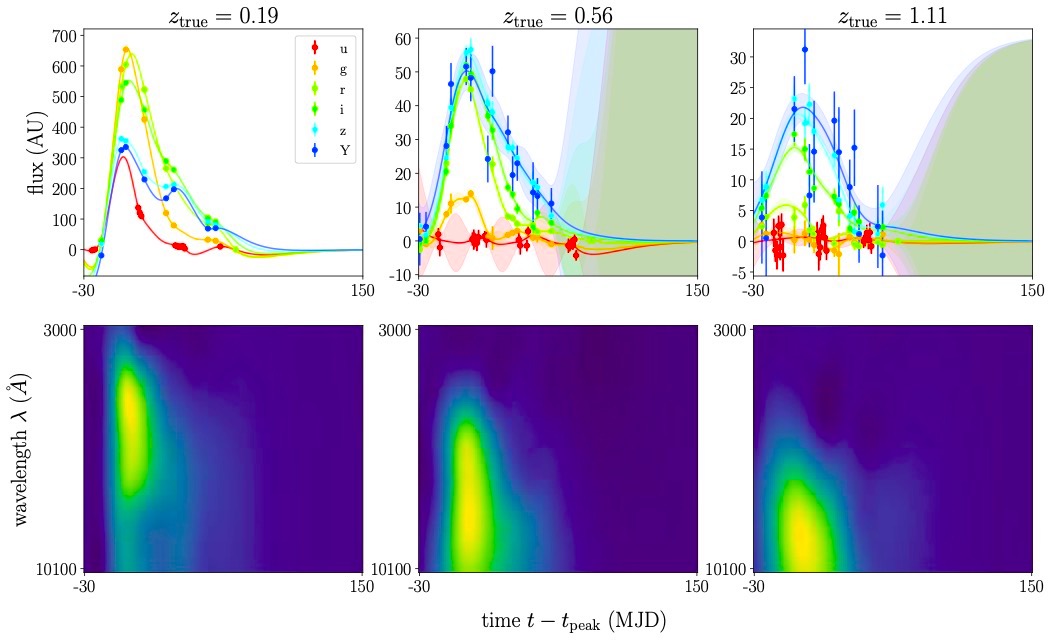}
    \caption{Examples of a low (left), medium (middle), and high (right) redshift simulated PLAsTiCC SN. \textbf{(top panel)} The multi-band lightcurves are shown in colored points, with Gaussian process predictions and uncertainties for this lightcurve shown in colored lines and shaded regions, respectively. \textbf{(bottom panel)} The flux matrix created from the lightcurve in the top panel (see \S\ref{subsec:lc-preprocess} for details). Large flux values are colored in yellow while lower flux values are in dark blue. The physical effects of redshifting, including longer wavelengths and longer transient durations, are observed in the flux matrices in the downward (towards higher wavelengths) shift of the yellow high flux region and the increased width of the yellow region with increasing redshift.}
    \label{fig:plasticc-lc-examples}
\end{figure*}
\subsection{Data Preprocessing}
\label{subsec:preprocess}

All SNe in all datasets are fit with a $\chi^2$-minimization program included in SNANA to determine several restframe parameters under the assumption that the event is a SN Ia: the time of SN peak brightness $t_\mathrm{peak}$, a stretch-like parameter $x_1$, a color parameter $c$, and the lightcurve normalization parameter $x_0$, as well as their uncertainties and covariances (i.e., $\sigma_{x_1}$, etc.). These parameters are used to calculate the distance modulus $\mu$, allowing the SNe to be placed on the Hubble diagram. SDSS datasets are fit with the SALT3 model while the LSST dataset is fit with the same extended SALT2 model used for the simulations.

We additionally select only SNe in our datasets that are well-described by the SALT model. These criteria were chosen following those used in past cosmology analyses, e.g. \citet{jla}:

\begin{itemize}
    \item $|c| < 0.3$
    \item $|x_1| < 3$
    \item $\sigma_{x_1} \leq 1$
    \item $\sigma_{t_\mathrm{peak}} \leq 2$
\end{itemize}

Finally, we choose SNe within the redshift range $0.01 \leq z \leq 0.4$ for SDSS and $0.01 \leq z \leq 1.2$ for LSST that have a successful LCFIT+Z \citep{Kessler_2010} fit, in order to draw direct comparisons between the performance of \name\ and LCFIT+Z. A more detailed description of LCFIT+Z can be found in \S\ref{subsec:lcfit+z}.

146,069 simulated SDSS SNe Ia and 81,734 simulated LSST SNe Ia remain after all cuts, as well as 489 observed SDSS SNe photometrically classified as SNe Ia. The number of observed SDSS lightcurves remaining after each cut is described in Table~\ref{tbl:sdss-cuts}. The redshift distributions of the PLAsTiCC training and test datasets after selection cuts are shown in Figure~\ref{fig:plasticc-z-dist}, while the redshift distributions of the SDSS simulated and observed datasets are shown in Figure~\ref{fig:sdss-z-dist}. We note that, while most machine learning classifiers perform better when trained on a dataset with balanced classes (i.e. flat redshift distribution), we found that this was not the case for \name. While our simulations are able to generate any artificial redshift distribution, an unphysical one such as a flat distribution could lead to strange artifacts in the dataset, confusing the classifier.

\subsubsection{Lightcurve Preprocessing}
\label{subsec:lc-preprocess}
To preprocess each lightcurve, we first take observations within the range $t_\mathrm{peak}-30 \leq t \leq t_\mathrm{peak}+150$, where $t_\mathrm{peak}$ is the epoch of SN peak brightness estimated by SNANA. This is to ensure the lightcurves contain just the SN and no extraneous information.

We use 2-dimensional Gaussian process (GP) regression to model each SN lightcurve in time ($t$) and wavelength ($\lambda$) space, then use the predictions of the fitted GP on a fixed $(\lambda, t)$ grid as the input to our neural network model. Modelling lightcurves of astronomical transients using 2D Gaussian processes was originally introduced in \citet{Boone_2019}; and \citet[][Q21]{scone} used the GP models to create ``images" from lightcurve data, which is the technique used for this work. Following Q21, we use the Mat\'{e}rn kernel ($\nu=3/2$) with a fixed $6000$\AA\ length scale in wavelength space and fit for the time length scale as well as the amplitude using maximum likelihood estimation. This GP model is fit separately to each lightcurve and used to produce a smooth 2D representation of the lightcurve by predicting flux values at each point in a $(\lambda, t)$ grid. We choose 32 equally spaced points in the range $3,000$\AA\ $\leq \lambda \leq 10,100$\AA\ ($\delta \lambda = 221.875$\AA) and 180 points in the range $t_\mathrm{peak}-30 \leq t \leq t_\mathrm{peak}+150$ ($\delta t=1$ day). This produces a $32 \times 180$ matrix of predicted flux values. We also produce a matrix of prediction uncertainties at each $\lambda_i, t_j$ of equal size.

We stack the flux and uncertainty matrices depthwise to produce a $32 \times 180 \times 2$ tensor and divide elementwise by the maximum flux value to constrain all entries to [0,1].

We show low, medium, and high redshift examples of PLAsTiCC lightcurves, the fitted GP models and their uncertainties, and the resulting flux matrices in Figure~\ref{fig:plasticc-lc-examples}. Redshifting increases the wavelength of light, which we see in the figure as the yellow (high flux) region moving down (toward $\lambda = 10,100$\AA) with increasing redshift. We also expect to observe a longer duration for higher redshift transients, which is evident in the increase in width of the yellow region. This data format is particularly well suited to the redshift prediction task, as we are able to visibly see expected physical results of redshifting in the flux matrices.

\subsubsection{Redshift Preprocessing}
The redshift range for each survey is discretized into $n_z$ discrete and non-overlapping bins ($0.01 \leq z \leq 0.4, n_z = 50$ for SDSS and $0.01 \leq z \leq 1.2, n_z = 150$ for LSST). We chose these $n_z$ values to preserve the approximate width of each redshift bin across surveys ($\delta z \sim 0.0078$). The bin corresponding to the true redshift of each SN is one-hot encoded and passed into the model as the training label.

\section{Model}
\name\ is a convolutional neural network model \citep[e.g. ][]{lecun,  zeiler2014visualizing, Simonyan2014VeryDC, krizhevsky2017imagenet} that takes in GP-interpolated lightcurves as well as the GP prediction uncertainties, prepared as described in \S\ref{subsec:preprocess}, and predicts a probability distribution over fine-grained redshift bins. This approach allows us to produce a discretized full PDF over redshift space for each individual SN without any assumptions on the underlying distribution, and has been used in a variety of contexts including image generation \citep{pixelrnn} and prediction of precipitation probabilities \citep{metnet}. Treating this as a categorical classification problem using the cross-entropy loss function has also been shown to accurately approximate Bayesian posterior probabilities \citep{lippman}. 

\subsection{Convolutional Neural Networks}
The convolutional neural network (CNN) is a class of artificial neural network with properties particularly suited to object and image recognition. It requires fewer trainable parameters than the standard feedforward network due to the convolution operation, learning a single weight matrix for small neighborhoods of the input image. This property is not only parameter-efficient but also imparts CNNs with translation-equivariance, i.e. the same feature shifted by $n$ pixels will produce the same response shifted by $n$ pixels. These convolutional layers are paired with pooling layers, which downsample the input to allow for the next set of convolutional layers to learn hierarchically more complex features. CNNs are prized in the machine learning community for being simple yet performant on image recognition benchmark datasets such as ImageNet \citep{imagenet}.

\subsection{Residual Learning}
\label{subsec:resnet}
We implement residual learning \citep{he2016deep} due to its state-of-the-art performance on the ImageNet benchmark at the time of publication as well as its widespread adoption over vanilla CNNs. Residual learning was presented as a solution to the degradation problem in CNNs, in which performance degraded past a certain threshold of network depth.  Since the additional layers could simply act as identity mappings and not affect the network performance, \citet{he2016deep} decreased the difficulty by not only feeding inputs sequentially through a series of layers, but also adding the inputs back in to the outputs of those layers. This allows the layers to learn the zero mapping rather than the identity mapping.
%
Since these layers are tasked with learning the \textit{residual} with respect to the input, this is known as \textit{residual learning} and the stack of layers is known as a \textit{residual block}. The residual connections in the \name\ architecture are shown as lines curving around each residual block in Figure~\ref{fig:architecture}.

\subsection{Architecture}
\label{subsec:architecture}
\begin{figure}
    \centering
    \includegraphics[scale=0.61, trim={16cm 11cm 9.5cm 26cm},clip]{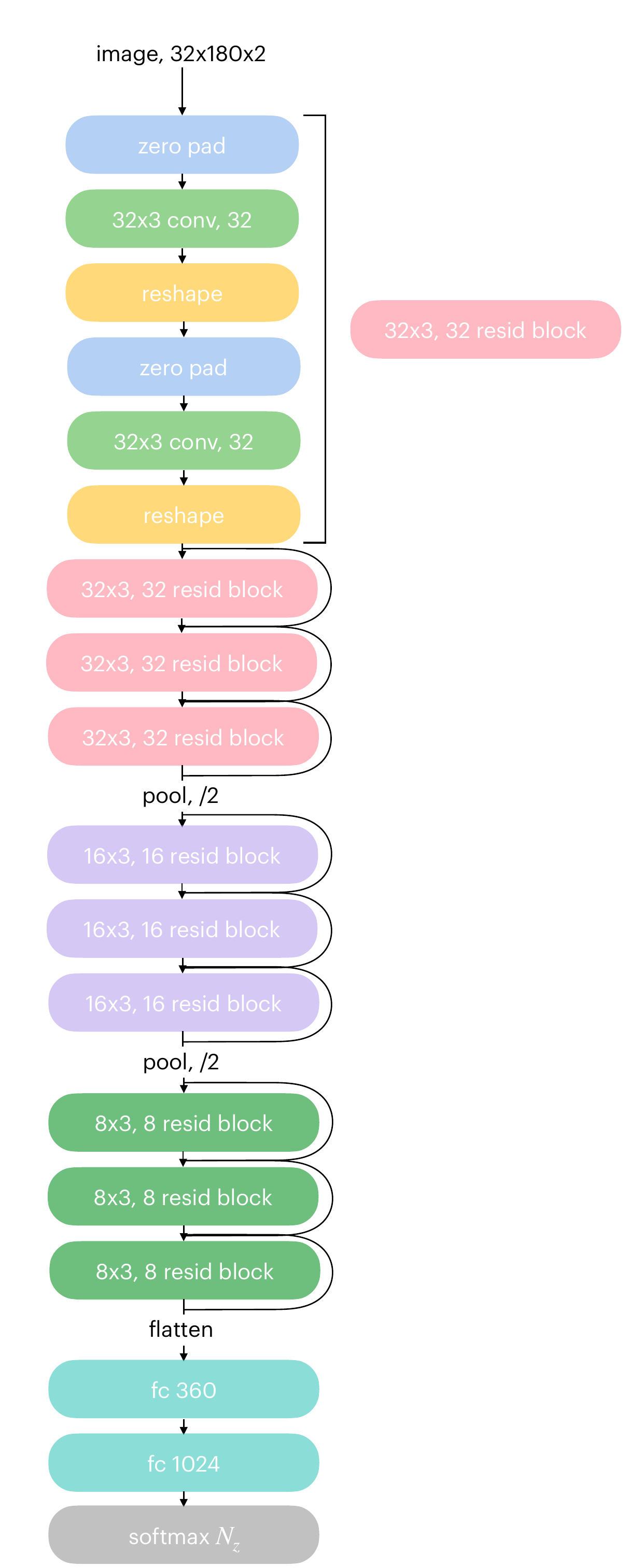}
    \caption{The model architecture developed for this work, described in the text in \S\ref{subsec:architecture}. A layer shown in the figure as ``$h \times d$ conv, $n_{\mathrm{filters}}$" is a 2D convolutional layer with a kernel size of $h \times d$ and $n_{\mathrm{filters}}$ filters. Similarly, ``$h \times d, n_{\mathrm{filters}}$ resid block" defines a residual block containing two $h \times d$ conv, $n_f$ convolutional layers. ``fc $n_{\mathrm{nodes}}$" denotes a fully connected layer with $n_{\mathrm{nodes}}$ nodes. The pooling layers apply max pooling, downsampling both the height and width by 2. The output softmax layer has $N_z$ nodes, $N_z=50$ for SDSS and $N_z=150$ for LSST.}
    \label{fig:architecture}
\end{figure}
Our model (Figure~\ref{fig:architecture}) takes as input a $32 \times 180 \times 2$ tensor containing the GP-interpolated lightcurve data on a $32 \times 180$ grid of wavelength and time values, and the GP uncertainties at each of those points. Table~\ref{tbl:layers} shows input and output dimensions of example layers in the model architecture. We also provide the model with the one-hot encoded vector specifying the correct redshift bin as the training label. 

\begin{table}
    \centering
    \begin{tabular}{l c c}
        \toprule
        Layer & Input Shape & Output Shape \\
        \midrule
        Zero Padding & $32 \times 180 \times 2$ & $32 \times 182 \times 2$ \\
        $32\times 3$ Convolutional & $32 \times 182 \times 2$ & $180 \times 32 \times 1$ \\
        Reshape & $180 \times 32 \times 1$ & $32 \times 180 \times 1$\\
        Max Pooling & $32 \times 180 \times 1$ & $16 \times 90 \times 1$\\
        \bottomrule
    \end{tabular}
    \caption{Description of example layers in the model architecture.}
    \label{tbl:layers}
\end{table}

The input is zero-padded on both sides to $32 \times 182 \times 2$ to ensure that feature maps output by the convolutional layers retain the original shape, then passed through a convolutional layer with 32 filters and kernel size $32 \times 3$. Since the convolutional kernel determines the receptive field of each unit in the layer, we choose a convolutional kernel that spans the wavelength space to allow each unit to learn from all wavelengths simultaneously while preserving the linearity of time. The output of this convolutional layer is $1 \times 180 \times 32$, which we then reshape back to $32 \times 180 \times 1$. This first feature map is now passed through a series of residual blocks.

Each residual block contains two convolutional blocks, each consisting of a ReLU nonlinearity, batch normalization, and the zero-padding, convolutional layer, and reshaping layer identical to the ones described above. The input to each residual block is then added to the output as described in \S\ref{subsec:resnet}. After each series of three residual blocks, the output is passed through a $2 \times 2$ max-pooling layer, downsampling the height and width of the output by a factor of 2.

Finally, the output is flattened and passed through a fully connected layer with 1,024 hidden nodes, which connects to the final softmax layer. The nodes in this layer correspond to redshift bins, thus it has 50 nodes for processing SDSS data and 150 for LSST data. The array of output probabilities is interpreted as the probability density over redshifts for our input SN.

\subsection{Calibration}

We also performed temperature scaling \citep{guo2017} to ensure that the probabilities output by \name\ are properly calibrated. In this process, we learn a single ``temperature" parameter used to scale the output probabilities. 

Before scaling, the output probabilities $\textbf{p}_i$ for input SN $i$ are derived from the softmax function 
\begin{equation}
    \textbf{p}_i^{(k)} = \sigma_{\text{SM}}(\textbf{q}_i)^{(k)} = \frac{\text{exp}(\textbf{q}_i^{(k)})}{\sum_{j=1}^{K} \text{exp}(\textbf{q}_i^{(j)})}
\end{equation}
where $\textbf{q}_i$ is the vector of network logits corresponding to SN $i$, i.e. the output of the final hidden layer of the network. The temperature parameter $T$ is learned by minimizing the cross-entropy loss between the one-hot encoded labels and the scaled probabilities, $\textbf{p}'_i$,
\begin{equation}
\textbf{p}'_i = \sigma_{\text{SM}}(\textbf{q}_i / T)    
\end{equation}

\begin{figure}
    \centering
    \includegraphics[scale=0.6, trim={0 0 0 0},clip]{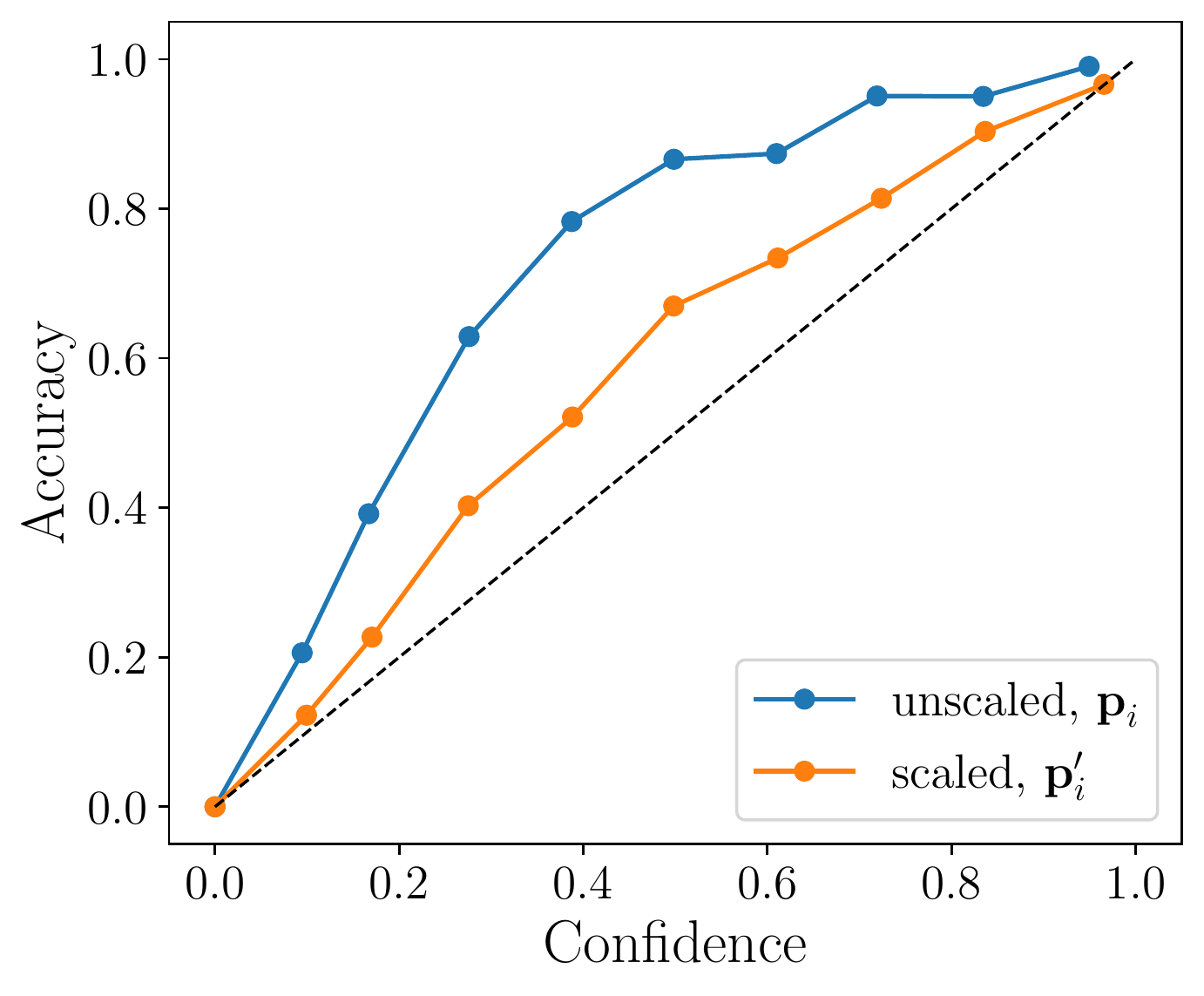}
    \caption{Reliability diagram showing PLAsTiCC model calibration before and after temperature scaling.}
    \label{fig:calibration}
\end{figure}

Calibration is typically evaluated using reliability diagrams \citep{degroot, niculescu} and the expected calibration error statistic \citep[ECE,][]{naeini}. Reliability diagrams show the prediction accuracy as a function of \textit{confidence}, which is defined as the probability associated with the predicted class: $\text{max}(\textbf{p}_i)$. A reliability diagram for a perfectly calibrated classifier will show the identity function, and any deviation from a perfect diagonal is a sign of miscalibration. The reliability diagram before and after temperature scaling for the PLAsTiCC model is shown in Figure~\ref{fig:calibration}. The scaled probabilities are much closer to the diagonal, representing a significant improvement in calibration. 

ECE is a weighted average of the difference between the accuracy and the confidence in bins of confidence values. ECE is more precisely defined as
\begin{equation}
    \text{ECE} = \sum_{m=1}^M \frac{|B_m|}{N} |\text{acc}(B_m) - \text{conf}(B_m)|
\end{equation}
where $B_m$ is the $m^{\text{th}}$ confidence bin and $N$ is the total number of samples in the dataset.
Prior to temperature scaling, the ECE for the PLAsTiCC model probabilities was 0.24, which improved to 0.08 after scaling.

\subsection{Implementation Details}
We split both the SDSS and LSST datasets into 90\% training, 5\% validation, and 5\% test datasets and trained both models with batch sizes of 2048 for 750 epochs. We minimize the cross-entropy loss function with the Adam optimizer \citep{adam} with an initial learning rate of 1e-3 that is halved after 25 epochs of no improvement in the validation loss. The total number of trainable parameters in the SDSS model is 451,654 and 554,154 for the LSST model due to the difference in number of redshift bins and resulting difference in output layer size. To prevent overfitting, we use a weight decay of 1e-3 as well as dropout layers \citep{dropout}. Calibration was performed with the validation set of both datasets and with an initialization of $T=1$. We minimize a cross-entropy loss function using Adam and find $T_{\text{SDSS}}=0.82$ and $T_{\text{LSST}}=0.64$.

\begin{figure*}
    \centering
    \includegraphics[scale=0.55]{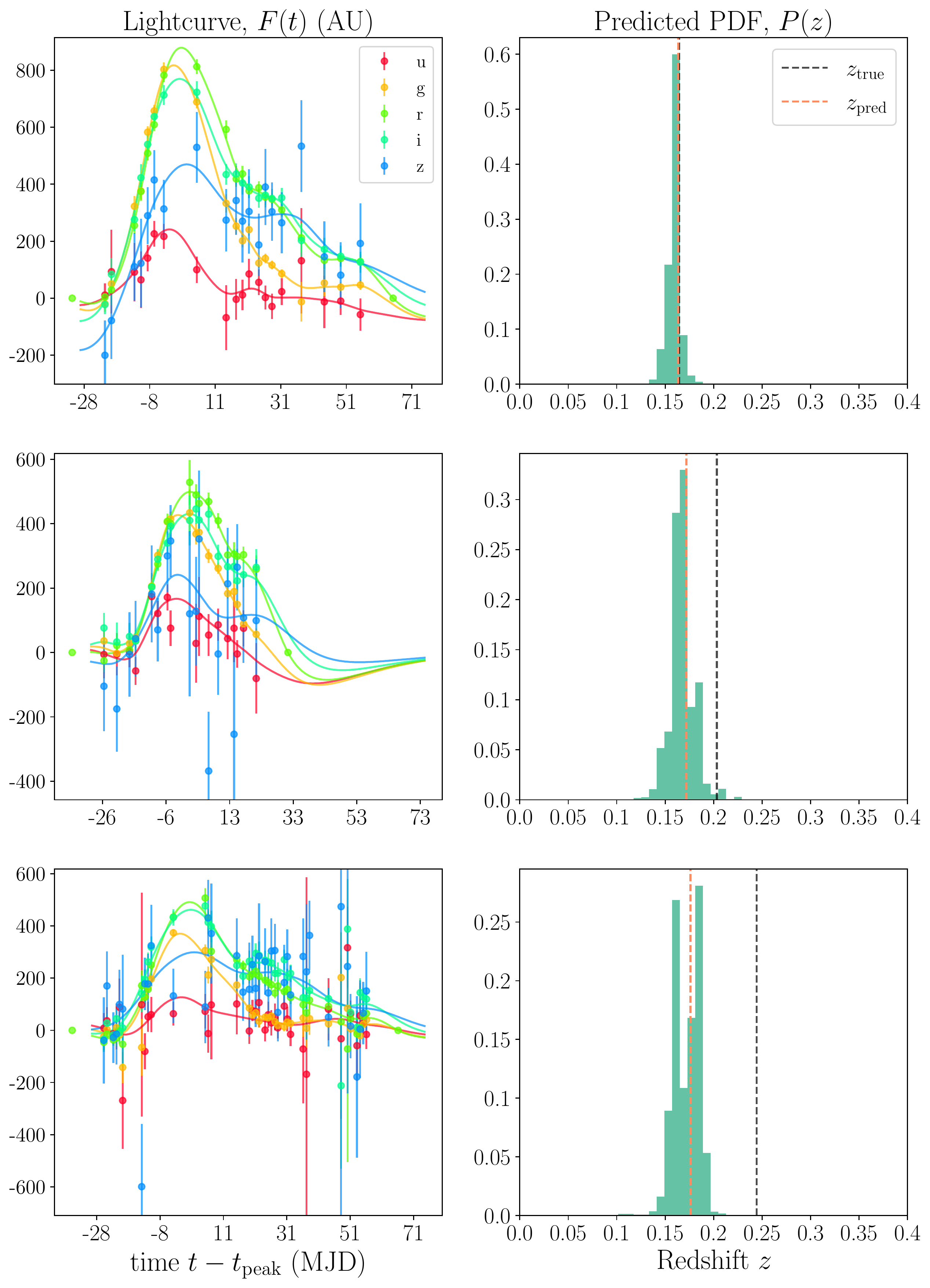}
    \caption{Examples of a high accuracy ($\Delta z < 0.005$, top), medium accuracy ($0.01 \leq \Delta z \leq 0.05$, middle), and outlier ($\Delta z > 0.05$, bottom) lightcurve and Gaussian process fit from the SDSS spectroscopic sample and their predicted PDFs. The true (spectroscopic) redshift is shown in the black dotted line, while the predicted redshift (mean(PDF)) is shown in orange. The medium and poor accuracy lightcurves have larger errors and scatter than the high accuracy lightcurve, and the low accuracy lightcurve appears to have much fewer points.}
    \label{fig:sdss-lc-examples}
\end{figure*}
 
\section{Results and Discussion}
Examples of lightcurves and their corresponding predicted redshift PDFs are shown in Figure~\ref{fig:sdss-lc-examples}. We chose spectroscopically confirmed SDSS SNe Ia lightcurves, and selected a high accuracy ($\Delta z < 0.005, \Delta z \equiv \frac{z_{\mathrm{pred}}-z_{\mathrm{true}}}{1+z_{\mathrm{true}}}$), medium accuracy ($0.01 \leq \Delta z \leq 0.05$), and an outlier ($\Delta z > 0.05$) example.

\subsection{Evaluation Metrics and Basis for Comparison}
\subsubsection{Point Estimates and Metrics}
\label{subsec:metrics}
\begin{figure}
    \centering
    \includegraphics[scale=0.45]{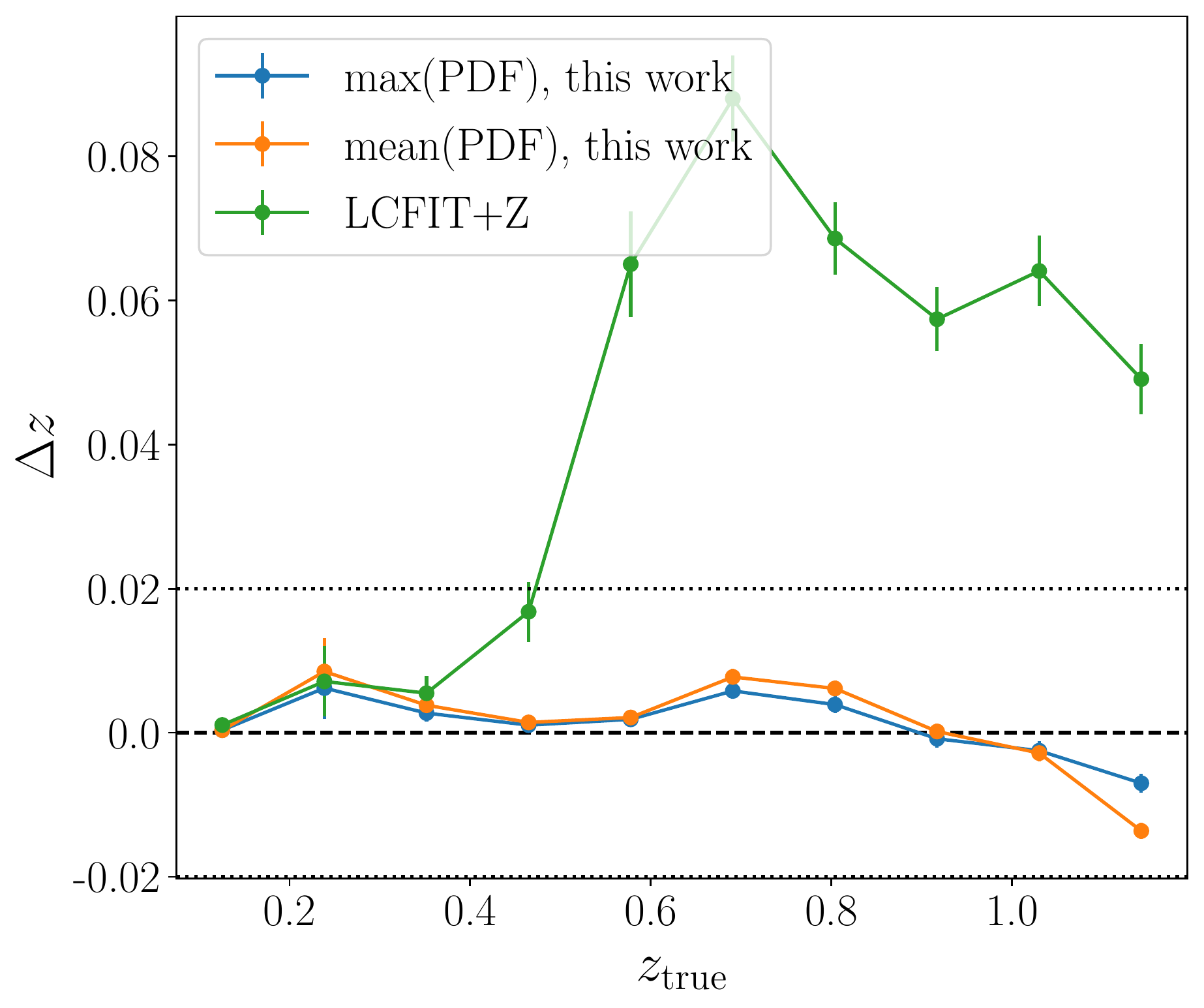}
    \caption{Mean binned residuals $\Delta z$ as a function of true redshift $z_{\mathrm{true}}$ for the PLAsTiCC simulated SNe Ia test set. Predictions from our model have much lower biases as well as scatter compared to predictions from LCFIT+Z. The max(PDF) and mean(PDF) point estimates for our model also agree quite well, resulting in similar $\Delta z$ values. Dotted lines are plotted at $\Delta z = \pm 0.02$ for reference.}
    \label{fig:plasticc-resids}
\end{figure}

\begin{figure*}
    \centering
    \includegraphics[scale=0.55]{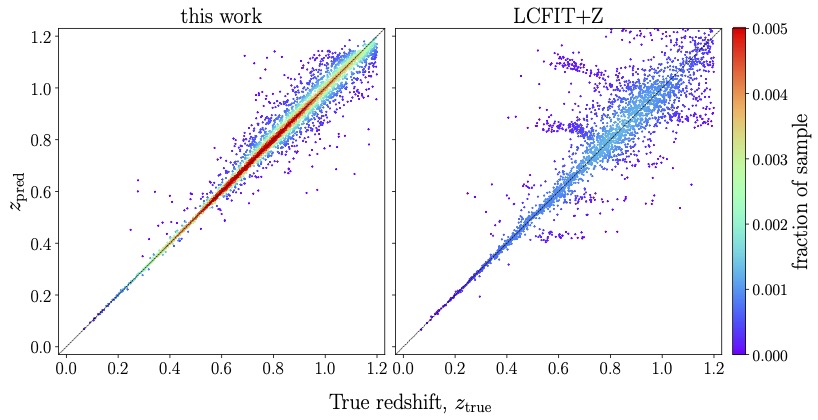}
    \caption{Predicted vs. true redshifts for the PLAsTiCC simulated SNe Ia sample colored by the fraction of each sample represented by each point. \textbf{(left)} Predictions from \name\ described in this work, \textbf{(right)} predictions from LCFIT+Z.}
    \label{fig:plasticc-scatter}
\end{figure*}

Although full photometric redshift PDFs are preferred for further statistical analyses (i.e. cosmological analyses), point estimates can also be computed from each PDF. We require these point estimates to evaluate our model performance against the true redshift values. We compare two possible methods to condense a PDF into a point estimate:
\begin{equation}
     \text{mean(PDF)} = \frac{1}{2} \sum_i p(Z_i) \frac{\lceil Z_{i} \rceil ^2 - \lfloor Z_{i} \rfloor ^2}{\lceil Z_{i} \rceil - \lfloor Z_{i} \rfloor}
\end{equation}
where $Z$ represents the vector of redshift bins, $\lceil Z_i \rceil $ the right edge of bin $i$, $\lfloor Z_i \rfloor $ the left edge of bin $i$, and $p(Z_i)$ the output probability assigned by the model to bin $i$; and
\begin{equation}
    \text{max(PDF)} = \tilde{Z}_{\text{argmax}_i(p(Z_i))}\\
\end{equation}
where $\tilde{Z}$ represents the array of midpoints of the redshift bins $Z$. Taking the weighted mean is a common summary statistic for PDFs, while taking the bin with maximum probability as the predicted output is typical for classification tasks. As shown in Tables~\ref{tab:plasticc-results} and~\ref{tab:sdss-results} and Figures~\ref{fig:plasticc-resids} and~\ref{fig:sdss-delta-z}, the two methods give similar results with mean(PDF) performing slightly better on the real SDSS dataset, and thus we use the mean point estimate for Figures~\ref{fig:plasticc-scatter},~\ref{fig:sdss-scatter},~\ref{fig:sdss-delta-z-corrected},~\ref{fig:hubble}, and~\ref{fig:sdss-delta-z-prior}.

We compute the following metrics for both of our point estimates, following e.g. \citet{Pasquet_2018}:
\begin{itemize}
    \item the residuals $\Delta z \equiv \frac{z_{\mathrm{pred}}-z_{\mathrm{true}}}{1+z_{\mathrm{true}}}$,
    \item the bias $\langle \Delta z \rangle$,
    \item the mean absolute deviation $\sigma_{\mathrm{MAD}} = 1.4826 \times \text{median}(|\Delta z - \text{median}(\Delta z)|)$,
    \item the fraction of outliers $\eta$ with $|\Delta z|>0.05$.
\end{itemize}

We evaluate \name\ using these metrics on the PLAsTiCC dataset as well as the simulated and observed SDSS datasets, and show the results in Tables~\ref{tab:plasticc-results} and~\ref{tab:sdss-results}.

\subsubsection{Comparison with LCFIT+Z}
\label{subsec:lcfit+z}
LCFIT+Z \citep{Kessler_2010} is an extension of the lightcurve fitting code described in \S\ref{subsec:data} that treats redshift as an additional free parameter. It determines the best fit lightcurve parameters by minimizing $\chi^2$ values of the observed and model fluxes, which are computed as a function of the free parameters (4 lightcurve parameters $t_{\mathrm{peak}}$, stretch $x_1$, color $c$, flux normalization $x_0$; and redshift $z_{\mathrm{phot}}$). LCFIT+Z is actively used for recent and ongoing experiments performing full cosmological analyses with photometric redshifts \citep[e.g.][]{Dai_2018, mitra2022}. LCFIT+Z produces point estimates with uncertainties as opposed to a full PDF, so we perform comparisons using our point estimates. 

We compare our results with LCFIT+Z as opposed to other SN photometric redshift estimators \citep[e.g., ][]{wang2015,deOliveira:2022tvw} due to its demonstrated performance, widespread use, and integration into SNANA. We run LCFIT+Z on the same datasets used to evaluate our model, enabling a direct comparison. Note that most documented uses of LCFIT+Z place a prior on the redshift fit using the redshift of the SN host galaxy, resulting in much more constrained fits. However, since (1) host spectroscopic redshifts are unavailable in future survey environments, and (2) to draw direct comparisons with \name, which does not require any host galaxy information, we omit the host galaxy redshift prior when running LCFIT+Z. In order to compare how these methods perform using all available information in the photometric survey era, we test \name\ and LCFIT+Z with host galaxy \textit{photometric} redshift priors in Section~\ref{subsec:with-prior}.

\subsection{LSST (PLAsTiCC) Results}
\label{subsec:plasticc}

We evaluate our model and our baseline for comparison, LCFIT+Z, on a test set of 4,057 simulated PLAsTiCC-like lightcurves with true redshift distribution shown in the right panel of Figure~\ref{fig:plasticc-z-dist}. 
The values of the evaluation metrics for results from this work as well as LCFIT+Z are shown in Table~\ref{tab:plasticc-results}. The two methods for obtaining point estimates from \name\ PDFs, mean and max, give similar results, though the mean point estimates have a degraded $\sigma_{\mathrm{MAD}}$ value compared to the max point estimates but a smaller outlier rate. However, the differences between mean and max point estimates are orders of magnitude smaller than the improvement we see relative to LCFIT+Z. The mean gives a $\sim 3.2\times$ larger result than max on $\sigma_{\mathrm{MAD}}$, compared with a $\sim 180\times$ larger result from LCFIT+Z.

\begin{table}
    \centering
    \begin{tabular}{l c c c}
        \toprule
        Metric & \multicolumn{2}{c}{this work} & LCFIT+Z\\
        \cmidrule(lr){2-3}
        & mean & max & \\
        \midrule
         bias $\langle \Delta z \rangle$ & 0.00095 & \textbf{0.00075 }& 0.058\\
         $\sigma_{\mathrm{MAD}}$ & 0.0081 & \textbf{0.0025 }& 0.0450\\
         outlier rate $\eta$ & \textbf{3.87\% }& 4.24\% & 32.3\%\\
         \bottomrule
    \end{tabular}
    \caption{Evaluation metrics computed for the PLAsTiCC test dataset for both the mean(PDF) and max(PDF) point estimates for our model as well as LCFIT+Z. The best result for each metric is shown in bold.}
    \label{tab:plasticc-results}
\end{table}

We show the mean binned residuals, $\Delta z$, as a function of true redshift for our model and LCFIT+Z in Figure~\ref{fig:plasticc-resids}. We see that while the residuals of our model and LCFIT+Z match quite well up to $z_{\mathrm{true}} \sim 0.4$, LCFIT+Z has a tendency to dramatically overestimate at higher redshifts. We also see that the spread within each residual bin in LCFIT+Z results is much larger, as shown in the size of the error bars on each point. In contrast, our model shows a relatively flat $\Delta z$ over the full redshift range with minimal redshift-dependent bias, which has not been achieved by other SN photometric redshift estimators.

Figure~\ref{fig:plasticc-scatter} more clearly shows the spread of predicted redshifts, where $z_{\mathrm{pred}}$ is calculated as the mean(PDF) for our model. The predictions produced by our model (left panel) lie much closer to the $z_{\mathrm{pred}} = z_{\mathrm{true}}$ line with approximately equal amounts of scatter on either side, reinforcing the minimal redshift-dependent bias shown in Figure~\ref{fig:plasticc-resids}. The dark red area along the $z_{\mathrm{pred}} = z_{\mathrm{true}}$ line also indicates that most of the sample is localized there. The LCFIT+Z results (right panel) have much larger spread and little localization, as evidenced by the lack of red in the plot.

\subsection{SDSS Results}
\label{subsec:sdss}
\begin{table*}
    \centering
        \begin{tabular}{l c c c c c c}
        \toprule
        Metric & \multicolumn{3}{c}{SDSS simulated} & \multicolumn{3}{c}{SDSS real}\\
        \cmidrule(lr){2-4} \cmidrule(lr){5-7} 
         & this work (mean) & this work (max) & LCFIT+Z & this work (mean) & this work (max) &  LCFIT+Z \\
        \midrule
        bias $\langle \Delta z \rangle $ & \textbf{7.8e-5} & 0.00073 &  0.023 & \textbf{0.0050 }& 0.0052 & 0.027\\
         $\sigma_{\mathrm{MAD}}$ & 0.011 & \textbf{0.010 }& 0.028 & \textbf{0.018 }& 0.020 & 0.047 \\
         outlier rate $\eta$ & \textbf{0.85\% }& 1.16\% & 19.9\% & \textbf{5.14\%} & 5.78\% & 26.1\%\\ 
        \bottomrule
    \end{tabular}
         
    \caption{Evaluation metrics computed for the SDSS simulated and real test datasets for both the mean(PDF) and max(PDF) point estimates for our model as well as LCFIT+Z. The best result for each metric and dataset is shown in bold.}
    \label{tab:sdss-results}
\end{table*}
\begin{figure*}
    \centering
    \includegraphics[scale=0.5]{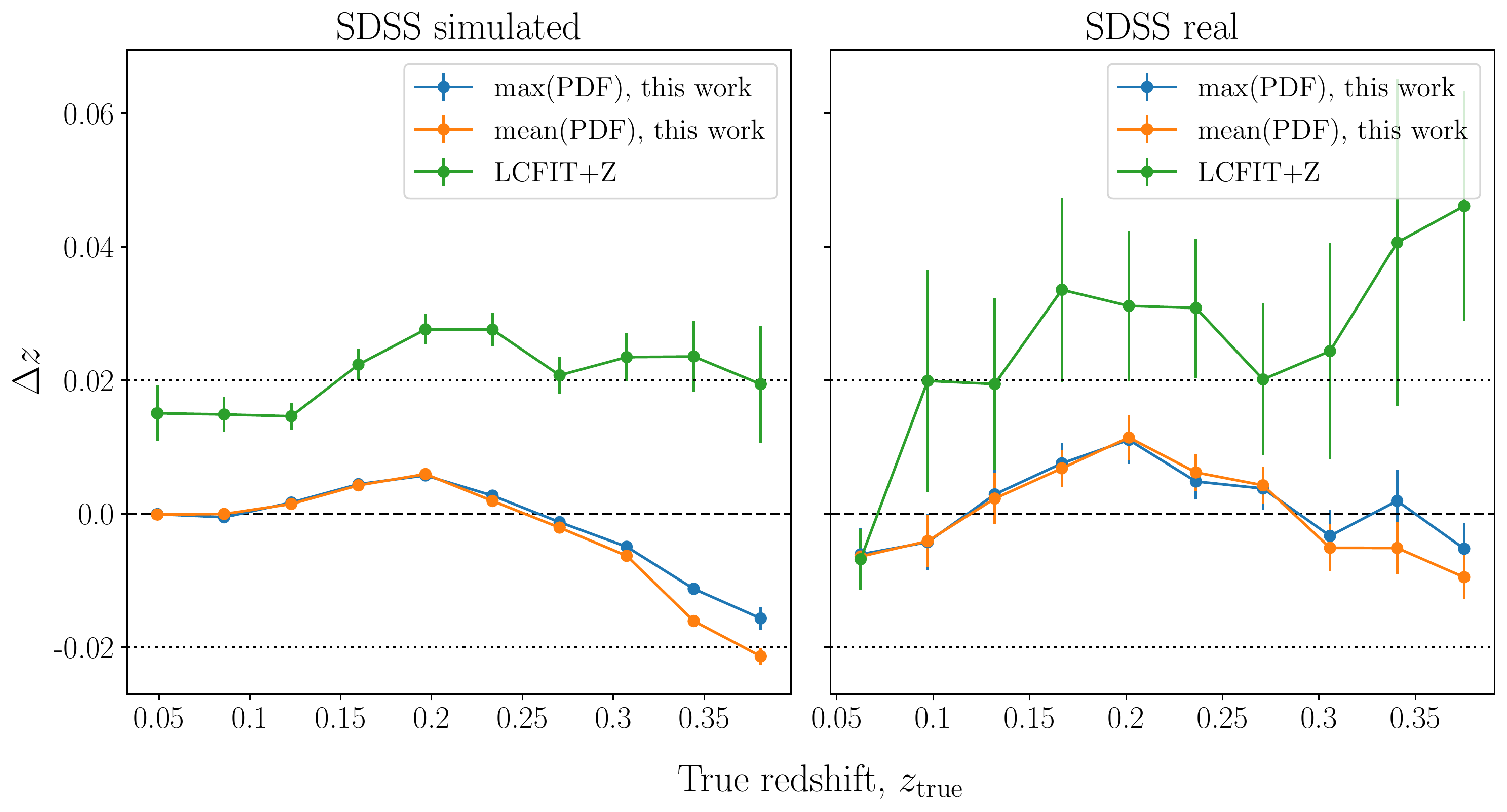}
    \caption{Mean binned residuals $\Delta z$ as a function of true redshift $z_{\mathrm{true}}$ for the SDSS simulated and real SNe Ia samples. Predictions from this work have much lower biases as well as scatter compared to predictions from LCFIT+Z. We also show that two common methods of condensing redshift PDFs into point estimates, max(PDF) and mean(PDF), resulting in similar errors for our model. Dotted lines are plotted at $\Delta z = \pm 0.02$ for reference.}
    \label{fig:sdss-delta-z}
\end{figure*}

\begin{figure*}
    \centering
    \includegraphics[scale=0.5]{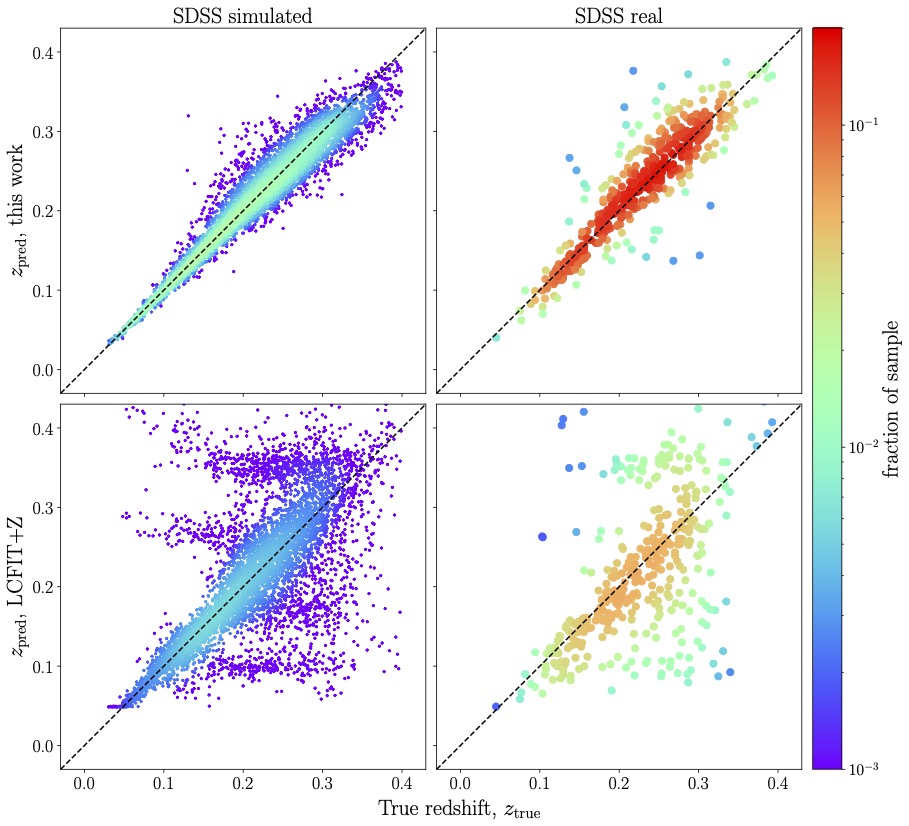}
    \caption{Predicted vs. true redshifts for the SDSS simulated and real SNe Ia samples colored by the fraction of each sample represented by each point. \textbf{(Top row)} Predictions from \name\ described in this work, \textbf{(bottom row}) predictions from LCFIT+Z. }
    \label{fig:sdss-scatter}
\end{figure*}

\begin{figure}
    \centering
    \includegraphics[scale=0.5]{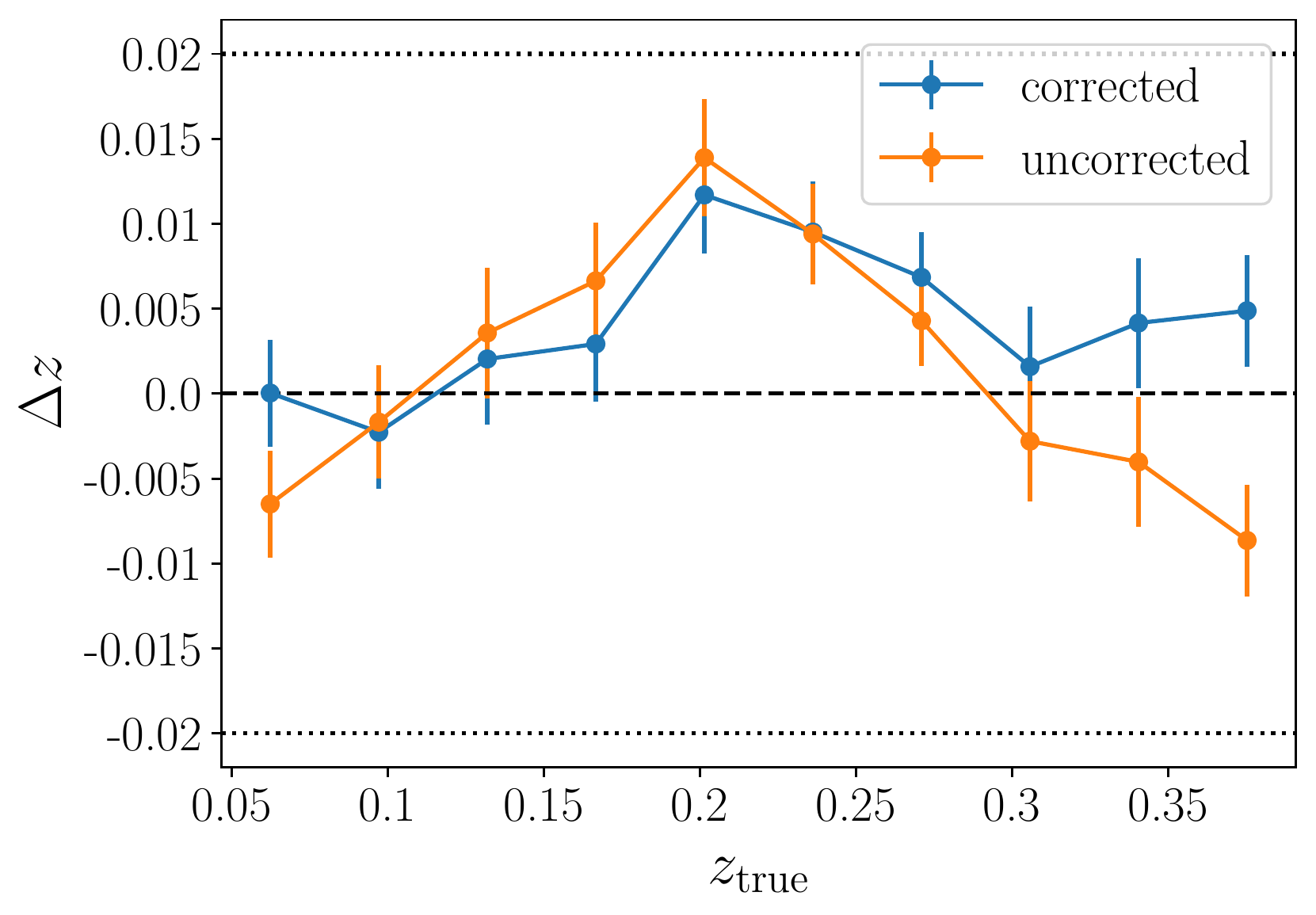}
    \caption{Predicted redshift error, $\Delta z$, as a function of true redshift, $z_{\mathrm{true}}$, for the SDSS real SNe Ia sample with and without bias correction computed from the simulated results. The point estimates used in this figure are computed using the mean(PDF) method. Dotted lines are plotted at $\Delta z = \pm 0.02$ for reference.}
    \label{fig:sdss-delta-z-corrected}
\end{figure}

We evaluate our model and LCFIT+Z on a test set of 5,274 simulated SDSS lightcurves and 489 real observed SDSS lightcurves with true redshift distributions shown in the right panel of Figure~\ref{fig:sdss-z-dist}. Table~\ref{tab:sdss-results} shows the evaluation metrics calculated with the mean and max point estimates on the simulated and observed SDSS datasets, as well as the LCFIT+Z point estimates. With the SDSS datasets, the similarity between the mean and max point estimates is more pronounced. We also note that though our model's performance degrades slightly between the simulated test set and the observed test set, all metrics still strongly favor our model as the better performer compared to LCFIT+Z. However, the $\sigma_{\mathrm{MAD}}$ values differ less between our model and LCFIT+Z compared to the PLAsTiCC test dataset, only offering a $2\times$ improvement as opposed to a $55-180\times$ improvement for the PLAsTiCC test dataset. This could be due to the narrower redshift range of SDSS, as the LCFIT+Z performance on the PLAsTiCC dataset degrades significantly after $z_{\mathrm{true}} \sim 0.4$. 

The mean binned residuals for our model and LCFIT+Z evaluated on the simulated and real SDSS datasets are shown in Figure~\ref{fig:sdss-delta-z}. The residuals from both models are much more constrained for SDSS than PLAsTiCC, as expected from the less significant differences in evaluation metrics. LCFIT+Z still exhibits larger mean $\Delta z$ values for both the simulated and real datasets as well as a larger spread in each bin. Both models generalize relatively well from simulations to real data, with our mean residuals staying within $|\Delta z| < 0.02$ and LCFIT within $|\Delta z| < 0.05$; however, an overall increase in $|\Delta z|$ values is noticeable between simulated and real data.

We show the comparison between predicted and true redshifts in Figure~\ref{fig:sdss-scatter} for the same datasets used in Figure~\ref{fig:sdss-delta-z}. Our model (top row) clearly produces more constrained predictions, as they lie much closer to the $z_{\mathrm{true}}=z_{\mathrm{pred}}$ line with a high density of points (shown in red) along the line. In contrast with the PLAsTiCC results, in which LCFIT+Z performed relatively well at lower redshifts, the performance over the full SDSS redshift range is poor. 

\subsubsection{Bias Correction}

Simulations not only allow deep learning methods such as \name\ to train on large datasets that would be infeasible with real data alone, but also give valuable estimates of the biases produced from those models that can be used to correct the results on real data. Here, we test this bias correction method on the results of \name\ on the SDSS simulated and real data. We compute the average $\Delta z$ values for the simulated SDSS sample in 10 bins of $z_{\mathrm{true}}$ values and bin the real SDSS sample in the same way. We then subtract the simulated $\Delta z$ value associated with the bin of each real $z_{\mathrm{pred}}$ estimate to produce the corrected curve in Figure~\ref{fig:sdss-delta-z-corrected}. Specifically, the corrected value of the $i^{\text{th}}$ SN in the real SDSS dataset belonging to bin $m$ is computed as 
\begin{equation}
z_{\text{corrected},i} = z_i - \langle \Delta z_{\text{sim},m} \rangle
\end{equation}
where $z_{\text{corrected},i}$ is the corrected prediction, $z_i$ is the uncorrected prediction, and $\langle \Delta z_{\text{sim},m} \rangle $ is the average $\Delta z$ from the simulated SNe in bin $m$.
The corrected $\Delta z$ values are computed as defined in \S\ref{subsec:metrics},
\begin{equation}
    \Delta z_{\text{corrected}} = \frac{z_{\text{corrected}}-z_{\text{true}}}{1+z_{\text{true}}}
\end{equation}
The corrected curve exhibits smaller biases than the uncorrected curve, showing that this bias correction method is valid in cases where the simulation is sufficiently representative of the real data.

\subsection{Further Experiments}
\subsubsection{Using \name\ photo-zs for Cosmology}
\begin{figure}
    \centering
    \includegraphics[scale=0.55]{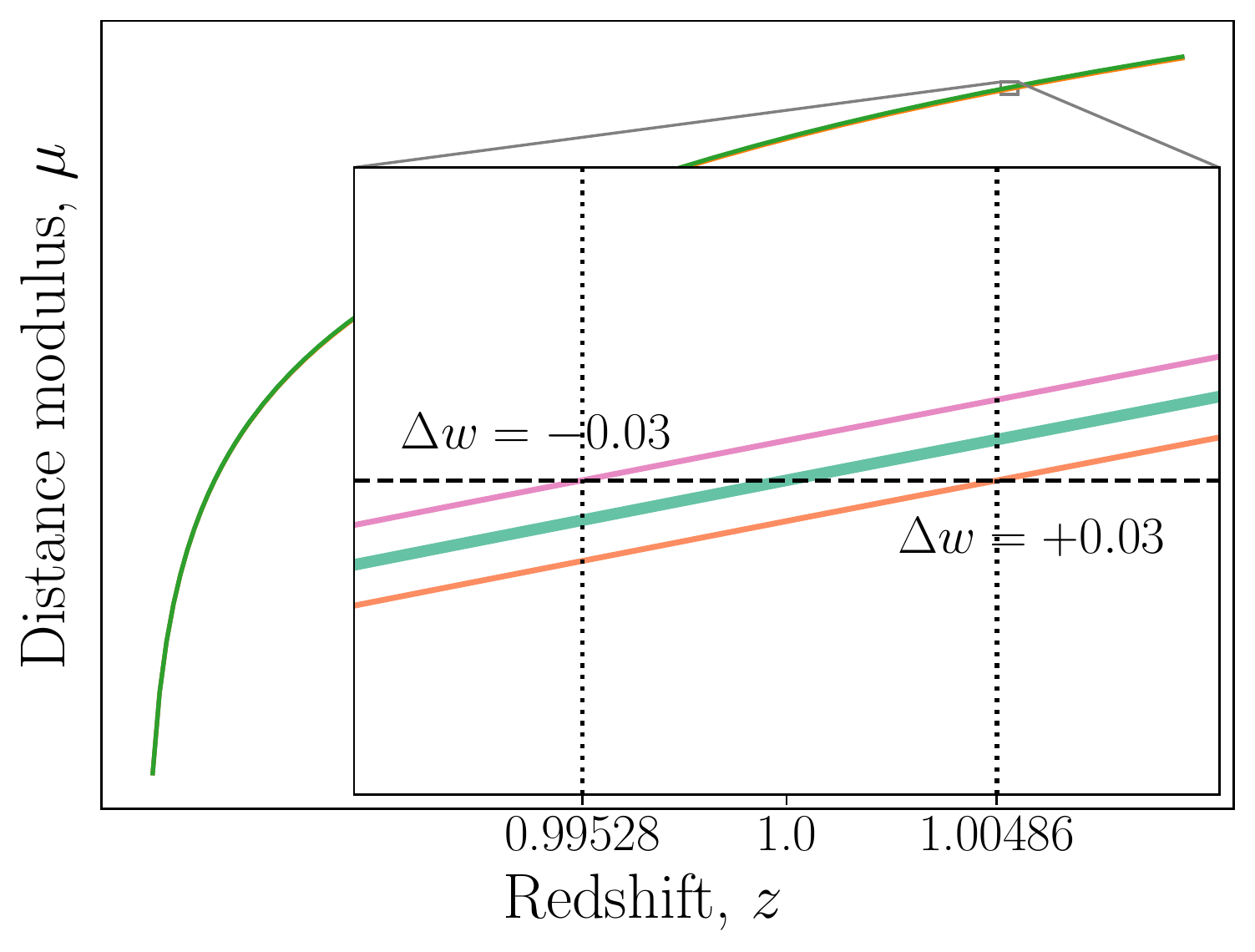}
    \includegraphics[scale=0.5]{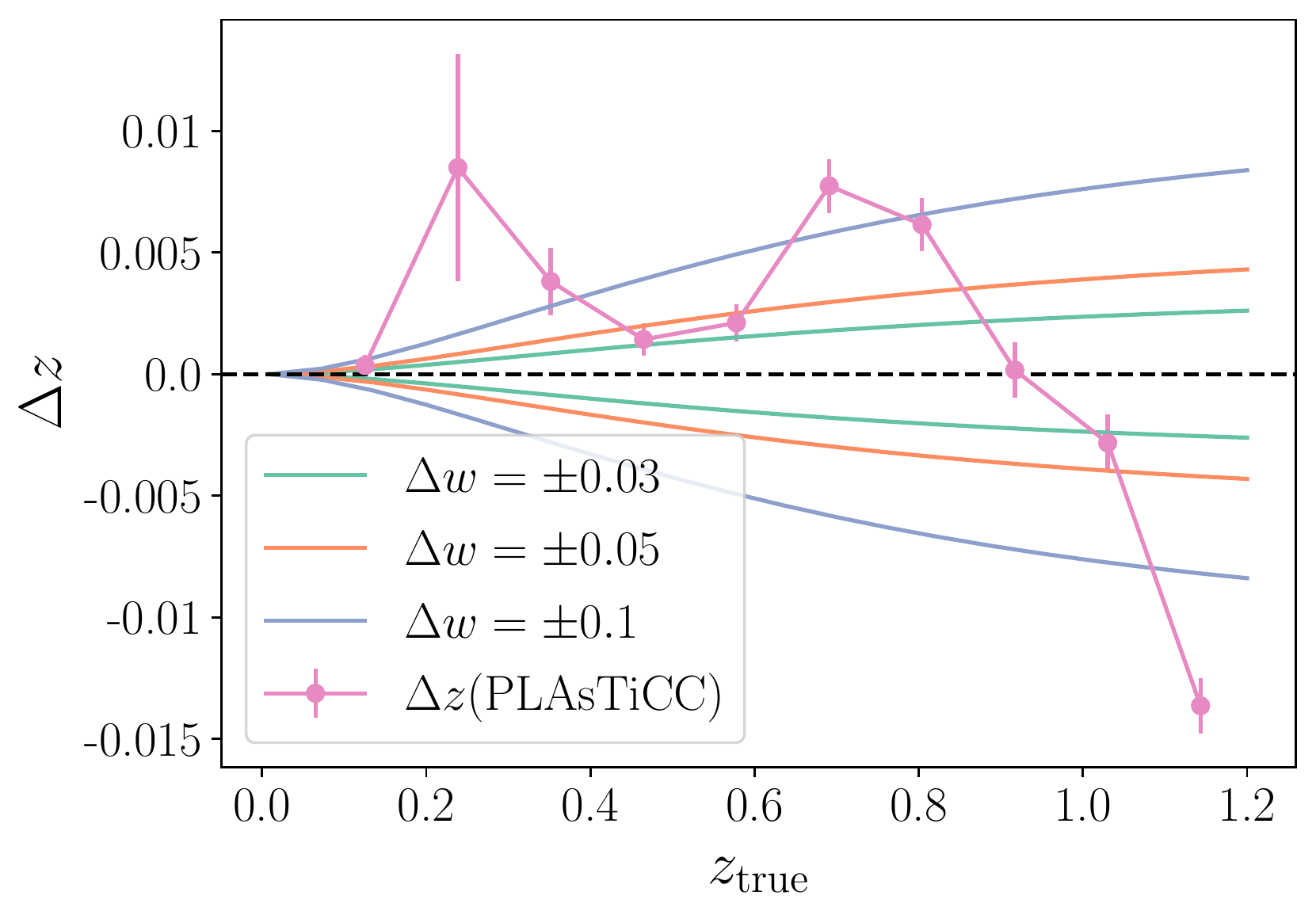}
    \caption{An illustration of our estimated cosmological biases arising from photo-$z$ errors. \textbf{(top)} A Hubble diagram zoomed in to the neighborhood of $z_{\text{model}}=1$ showing a fiducial cosmology (green: $w=-1, \Omega_m=0.3$) and two biased cosmologies (pink: $\Delta w = -0.03$, orange: $\Delta w = +0.03$). The dashed black line shows the distance modulus value at $z_{\text{model}}=1$ for the fiducial cosmology, $\mu_{\mathrm{model}}(z_{\text{model}}=1)$. The two dotted lines show the redshifts that correspond to the value of $\mu_{\mathrm{model}}(z_{\text{model}}=1)$ for the two biased cosmologies, i.e. $\mu_{\text{model}}(z_{\text{model}}=1) = \mu_{\Delta w = -0.03}(z=0.99528) = \mu_{\Delta w = +0.03}(z=1.00486)$. The values of these redshifts are labeled on the redshift axis. We approximate the redshift error required to create a bias of, e.g. $\Delta w = +0.03$, as the difference between the biased and fiducial redshift values, $d z_{\text{bias}} = 1-1.00486 = -0.00486$. \textbf{(bottom)} $\Delta z = dz / (1+z)$ values for various choices of $\Delta w$, compared to the mean binned residuals $\Delta z$ produced by the mean point estimates of \name\ PDFs (pink, reproduced from Figure~\ref{fig:plasticc-resids}).}
    \label{fig:hubble}
\end{figure}

Though producing a full cosmological analysis using photo-$z$s predicted by \name\ is outside the scope of this work, we provide some intuition for the quality of our estimates in the context of cosmology. The redshift error $dz_{\text{bias}}$ that results in a $w$ bias of $\Delta w$ at a particular redshift, $z_{\text{model}}$, can be approximated by the redshift difference 
\begin{equation}
    dz_{\text{bias}} = z_{\text{model}}-z_{\Delta w}
\end{equation} where $z_{\Delta w}$ is the redshift associated with the distance modulus $\mu_{\text{model}}(z_{\text{model}})$ for the biased cosmology, i.e.
\begin{equation}
    \mu_{\text{model}}(z_{\text{model}}) = \mu_{\Delta w}(z_{\Delta w})
\end{equation}
where $\mu_{\text{model}}, \mu_{\Delta w}$ are the distance moduli associated with the model and biased cosmologies, respectively.

A concrete example of this is illustrated in the top panel of Figure~\ref{fig:hubble}, which shows a zoomed-in portion of the Hubble diagram in the neighborhood of our chosen $z_{\text{model}}=1$ for our fiducial cosmology ($w=-1, \Omega_m=0.3$) as well as examples of biases on $w$ ($\Delta w \pm 0.03$) plotted on either side. Our choice of $z_{\text{model}}=1$ is motivated by the importance of high redshift SNe on the constraining power on $w$, so we compare with our PLAsTiCC results. We choose to focus on the $\Delta w = +0.03$ cosmology as our redshifts are slightly underestimated at $z=1$. In this example, $ z_{\Delta w = +0.03} = 1.00486$ at $z_{\text{model}} =1$ is shown as the dotted vertical line on the right.
We approximate $d z_{\text{bias}}$ for $\Delta w = +0.03$ at $z_{\text{model}} = 1$ to be \begin{equation}
    d z_{\text{bias}} = z_{\text{model}} - z_{\Delta w = +0.03} = 1 - 1.00486 = -0.00486
\end{equation} 
This corresponds to half the distance between the dotted vertical lines in the figure. The mean redshift error for our results on the PLAsTiCC dataset at $z_{\text{true}}=1$ are 
\begin{equation}
    \Delta z = \frac{z_{\text{pred}}-z_{\text{true}}}{1+z_{\text{true}}} = -0.002
\end{equation} translating to an expected redshift difference of
\begin{equation}
    z_{\text{pred}}-z_{\text{true}} = (1+z_{\text{true}}) \cdot -0.002 = -0.004
\end{equation}
This is below the expected $d z_{\text{bias}}$ associated with a $w$ shift of $\Delta w = +0.03$.

The bottom panel of Figure~\ref{fig:hubble} shows $\Delta z$ values for different choices of cosmological biases (here we choose to vary $w$) across the full LSST redshift range, along with the $\Delta z$ values produced by \name\ on the PLAsTiCC dataset. Note that the $y$ axis of this plot is $\Delta z = dz / (1+z)$, i.e. $dz_{\text{bias}}$ normalized by $(1+z_{\text{model}})$, for ease of comparison with \name\ mean residual $\Delta z$ values. This comparison over all redshifts shows that \name\ redshifts lack the precision to constrain cosmology to $\Delta w = \pm 0.03$, since the pink line representing \name\ redshift errors mostly does not lie in the area within the $\Delta w = \pm 0.03$ teal lines. However, this high-level analysis is an overestimate of the expected impact on cosmology, and a thorough cosmological analysis with \name\ photo-$z$'s will be the subject of a future work.

\subsubsection{Comparison with LCFIT+Z with Host Galaxy Redshift Prior}
\label{subsec:with-prior}

In \S\ref{subsec:plasticc} and \S\ref{subsec:sdss}, we showed results for \name\ and LCFIT+Z with no host galaxy information included. \name\ was formulated not to require a host redshift prior in order to prevent biases due to incorrect host redshifts. However, in a context in which host mismatches are rare and host redshifts are reliable, we want to use all available information to produce the most constrained SN photo-$z$ estimates.

We test both \name\ and LCFIT+Z with a prior on $z_{\text{pred}}$, the predicted redshift, given by the host galaxy photometric redshift estimate. We choose to use photometric redshifts as opposed to spectroscopic to more closely emulate a future scenario in which most host galaxies will not have spectroscopic information available.

We model this prior, $P(Z_{\text{host}})$, as a Gaussian centered on the photo-$z$ of the host galaxy, $z_{\text{host}}$, and use the estimated uncertainty, $\sigma_{z_{\text{host}}}$, as the standard deviation:
\begin{equation}
    Z_{\text{host}} \sim \mathcal{N}(z_{\text{host}}, \sigma_{z_{\text{host}}})
\end{equation}
We treat \name\ PDFs, $P(Z_{\text{Photo-$z$SN}})$, as Bayesian posteriors and apply the host prior using Bayes' theorem: 
\begin{equation}
\begin{aligned}
    P(Z_{\text{pred}}) = \frac{P(Z_{\text{Photo-$z$SN}} | Z_{\text{host}}) P(Z_{\text{host}})}{P(Z_{\text{Photo-$z$SN}})}\\
    = \frac{P(Z_{\text{Photo-$z$SN}})P(Z_{\text{host}})}{\sum_i P(Z_{\text{Photo-$z$SN},i})P(Z_{\text{host},i})}
\end{aligned}
\end{equation}
simplified by the fact that $P(Z_{\text{host}})$ and $P(Z_{\text{Photo-$z$SN}})$ are statistically independent.

LCFIT+Z uses a Markov chain Monte Carlo (MCMC) process to sample from the posterior distribution over a 5-dimensional parameter space: 4 SALT fit parameters color $c$, stretch-luminosity parameter $x_1$, time of peak brightness $t_0$, flux normalization parameter $x_0$; and redshift $z_{\text{phot}}$. The host prior constrains the search space, resulting in a higher likelihood of convergence to a global minimum.

\begin{figure}
    \centering
    \includegraphics[scale=0.47]{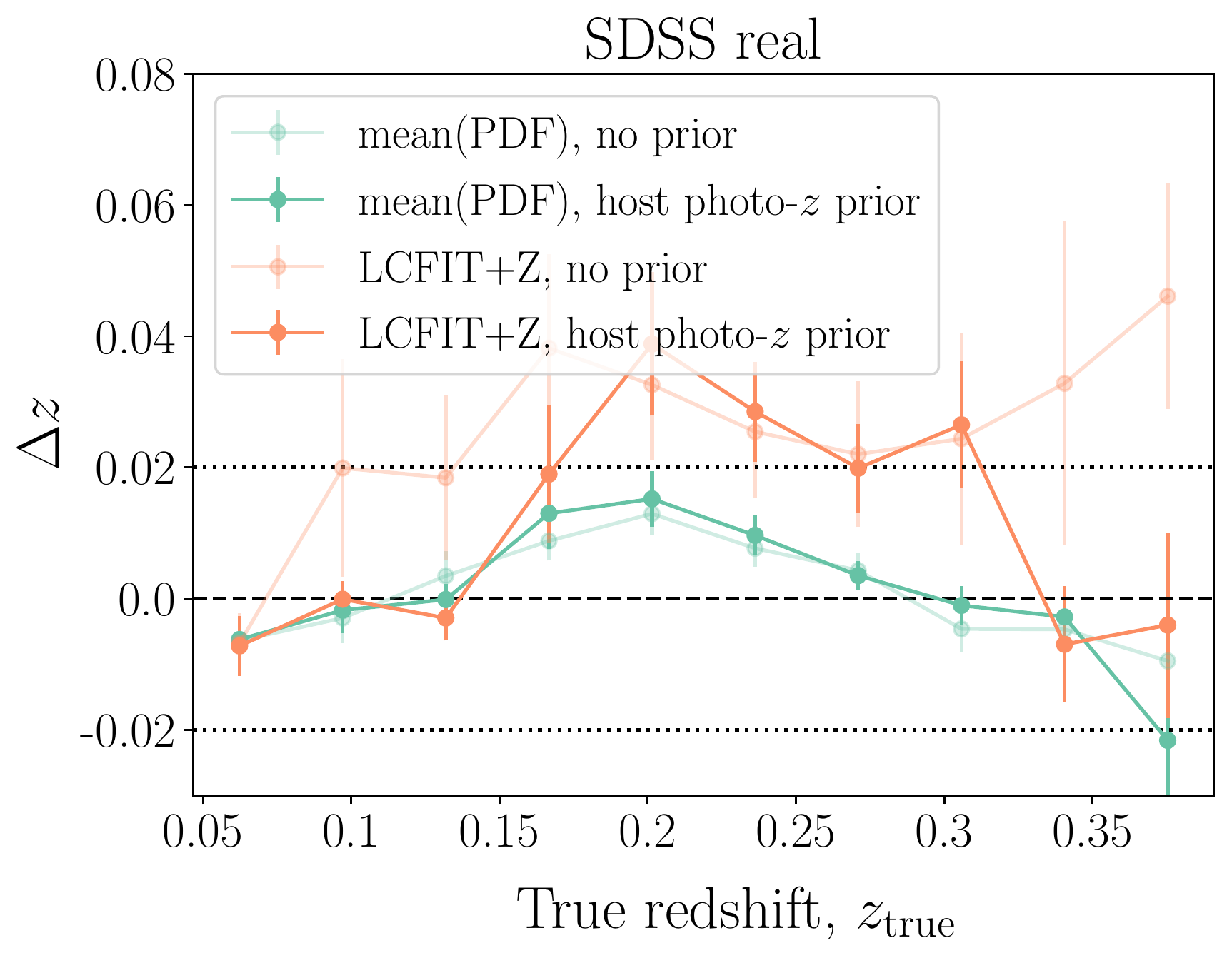}
    \caption{Mean binned residuals, $\Delta z \equiv \frac{z_{\mathrm{pred}} - z_{\mathrm{true}}}{1+z_{\mathrm{true}}}$, as a function of true redshift, $z_{\mathrm{true}}$, for the SDSS observed photometric SNe Ia sample. Errors from both \name\ and LCFIT+Z are shown with (darker) and without (lighter) a prior on $z_{\text{pred}}$ from the host galaxy photometric redshift. The lighter (no prior) curves are identical to those in Figure~\ref{fig:sdss-delta-z}, with the max(PDF) curves omitted for clarity.}
    \label{fig:sdss-delta-z-prior}
\end{figure}

In Figure~\ref{fig:sdss-delta-z-prior}, we show $\Delta z$ values for the SDSS photometric SNe Ia dataset for \name\ and LCFIT+Z with and without a host photo-$z$ prior. The $\Delta z$ results with no host prior are identical to those in the right panel of Figure~\ref{fig:sdss-delta-z}. The host prior noticeably improves LCFIT+Z results on $z_{\text{true}} < 0.2$ and $z_{\text{true}} > 0.3$, as well as the scatter across the full redshift range, shown in the smaller error bars. The \name\ results with and without host prior are very similar, however, as the host prior is often less constraining than the \name\ PDFs themselves. We show the bias, $\sigma_{\text{MAD}}$, and outlier rate for both models with and without the host prior in Table~\ref{tbl:sdss-prior}. These results show that \name\ outperforms LCFIT+Z even in the presence of a host galaxy redshift prior.

\begin{table}
    \centering
    \begin{tabular}{l c c c c}
        \toprule
        Metric & \multicolumn{2}{c}{this work} & \multicolumn{2}{c}{LCFIT+Z}\\
        \cmidrule(lr){2-3}  \cmidrule(lr){4-5}
        & no prior & with prior & no prior & with prior \\
        \midrule
         bias $\langle \Delta z \rangle$ & \textbf{0.0050} & 0.0061 & 0.027 & 0.021 \\
         $\sigma_{\mathrm{MAD}}$ & 0.018 & \textbf{0.017 }& 0.057 & 0.025\\
         outlier rate $\eta$ & 5.14\% & \textbf{4.93\% }& 26.1\% & 17.6\%\\
         \bottomrule
    \end{tabular}
    \caption{Evaluation metrics computed for the SDSS observed photometric SNe Ia sample both with and without a host galaxy photo-$z$ prior. The best results for each metric and dataset are shown in bold.}
    \label{tbl:sdss-prior}
\end{table}

\section{Conclusions}
In this work we presented \name, a convolutional neural network model that predicts full redshift PDFs from multi-band photometric SNe Ia lightcurves. We evaluated its performance on simulated SDSS and LSST lightcurves, as well as a photometrically confirmed SDSS SN Ia sample. We compared our results against LCFIT+Z, the most frequently used photometric redshift estimation method for SNe, and showed superior performance across all evaluation metrics. Our model also exhibits minimal redshift-dependent bias, which has plagued redshift estimators in the past, and generalizes well between simulated and observed data.

Though \name\ does not require host galaxy information to produce accurate SN photo-$z$ estimates, host galaxy redshifts can be incorporated as a prior to further improve our predictions. We tested \name\ and LCFIT+Z with a host galaxy photometric redshift prior to demonstrate the performance of these models in the LSST era when host galaxy spectroscopic redshifts will not be widely available. We show that while the host galaxy prior improves LCFIT+Z estimates, \name\ still produces more accurate photo-$z$ point estimates as well as constrained PDFs.

We envision \name\ to be useful for many tasks in supernova science, including precise volumetric rate calculations, discovery of incorrect host galaxy matches via redshift discrepancies, and photometric SN Ia cosmology. We briefly explored the cosmological constraints that can be expected from \name\ photo-$z$s and concluded that the bias on $w$ may be on the order of $\Delta w \sim 0.1$.

In future work, we intend to develop a framework for incorporating redshift PDFs and their associated uncertainties into the Hubble diagram and produce more accurately modeled cosmological constraints from \name\ redshift PDFs. We also plan to explore the use of \name\ redshift predictions as a method of identifying host confusion and potential mismatches for LSST. Though \name\ was developed for SNe Ia with cosmological applications in mind, the data processing and model architecture can be generalized for use with any astronomical time-domain events. 

Accurate photometric redshift estimation for SNe Ia will become vital in the imminent era of photometric SN Ia cosmology. We believe that the approach and model presented here will allow us to maximize the constraining power of these new datasets.

\section*{acknowledgments}
H.Q. and M.S. were supported by DOE grant DE-FOA-0002424 and NSF grant AST-2108094. The authors would like to thank Sang Michael Xie for insightful discussions on the network architecture and training process, Brodie Popovic for providing SDSS simulations, Rebecca Chen for helpful discussions and LCFIT+Z setup, Rick Kessler for assisting with LCFIT+Z, SNANA, and LSST simulations, and Carles S\'{a}nchez and Jaemyoung Lee for helpful discussions.

This research used resources of the National Energy Research Scientific Computing Center (NERSC), a U.S. Department of Energy Office of Science User Facility located at Lawrence Berkeley National Laboratory, operated under Contract No. DE-AC02-05CH11231. This work was completed in part with resources provided by the University of Chicago’s Research Computing Center.
\bibliography{references}

\begin{thebibliography}{}
\expandafter\ifx\csname natexlab\endcsname\relax\def\natexlab#1{#1}\fi
\providecommand{\url}[1]{\href{#1}{#1}}
\providecommand{\dodoi}[1]{doi:~\href{http://doi.org/#1}{\nolinkurl{#1}}}
\providecommand{\doeprint}[1]{\href{http://ascl.net/#1}{\nolinkurl{http://ascl.net/#1}}}
\providecommand{\doarXiv}[1]{\href{https://arxiv.org/abs/#1}{\nolinkurl{https://arxiv.org/abs/#1}}}

\bibitem[{Abbott {et~al.}(2019)Abbott, Allam, Andersen, Angus, Asorey, Avelino,
  Avila, Bassett, Bechtol, Bernstein, Bertin, Brooks, Brout, Brown, Burke,
  Calcino, Rosell, Carollo, Kind, Carretero, Casas, Castander, Cawthon,
  Challis, Childress, Clocchiatti, Cunha, D'Andrea, da~Costa, Davis, Davis,
  Vicente, DePoy, Desai, Diehl, Doel, Drlica-Wagner, Eifler, Evrard, Fernandez,
  Filippenko, Finley, Flaugher, Foley, Fosalba, Frieman, Galbany, Garc{\'{\i}
  }a-Bellido, Gaztanaga, Giannantonio, Glazebrook, Goldstein,
  Gonz{\'{a}}lez-Gait{\'{a}}n, Gruen, Gruendl, Gschwend, Gupta, Gutierrez,
  Hartley, Hinton, Hollowood, Honscheid, Hoormann, Hoyle, James, Jeltema,
  Johnson, Johnson, Kasai, Kent, Kessler, Kim, Kirshner, Kovacs, Krause, Kron,
  Kuehn, Kuhlmann, Kuropatkin, Lahav, Lasker, Lewis, Li, Lidman, Lima, Lin,
  Macaulay, Maia, Mandel, March, Marriner, Marshall, Martini, Menanteau,
  Miller, Miquel, Miranda, Mohr, Morganson, Muthukrishna, Möller, Neilsen,
  Nichol, Nord, Nugent, Ogando, Palmese, Pan, Plazas, Pursiainen, Romer,
  Roodman, Rozo, Rykoff, Sako, Sanchez, Scarpine, Schindler, Schubnell,
  Scolnic, Serrano, Sevilla-Noarbe, Sharp, Smith, Soares-Santos, Sobreira,
  Sommer, Spinka, Suchyta, Sullivan, Swann, Tarle, Thomas, Thomas, Troxel,
  Tucker, Uddin, Walker, Wester, Wiseman, Wolf, Yanny, Zhang, \&
  and}]{Abbott_2019}
Abbott, T. M.~C., Allam, S., Andersen, P., {et~al.} 2019, The Astrophysical
  Journal, 872, L30, \dodoi{10.3847/2041-8213/ab04fa}

\bibitem[{Benitez(2000)}]{Benitez_2000}
Benitez, N. 2000, The Astrophysical Journal, 536, 571, \dodoi{10.1086/308947}

\bibitem[{{Betoule} {et~al.}(2014){Betoule}, {Kessler}, {Guy}, {Mosher},
  {Hardin}, {Biswas}, {Astier}, {El-Hage}, {Konig}, {Kuhlmann}, {Marriner},
  {Pain}, {Regnault}, {Balland}, {Bassett}, {Brown}, {Campbell}, {Carlberg},
  {Cellier-Holzem}, {Cinabro}, {Conley}, {D'Andrea}, {DePoy}, {Doi}, {Ellis},
  {Fabbro}, {Filippenko}, {Foley}, {Frieman}, {Fouchez}, {Galbany}, {Goobar},
  {Gupta}, {Hill}, {Hlozek}, {Hogan}, {Hook}, {Howell}, {Jha}, {Le Guillou},
  {Leloudas}, {Lidman}, {Marshall}, {M{\"o}ller}, {Mour{\~a}o}, {Neveu},
  {Nichol}, {Olmstead}, {Palanque-Delabrouille}, {Perlmutter}, {Prieto},
  {Pritchet}, {Richmond}, {Riess}, {Ruhlmann-Kleider}, {Sako}, {Schahmaneche},
  {Schneider}, {Smith}, {Sollerman}, {Sullivan}, {Walton}, \& {Wheeler}}]{jla}
{Betoule}, M., {Kessler}, R., {Guy}, J., {et~al.} 2014, \aap, 568, A22,
  \dodoi{10.1051/0004-6361/201423413}

\bibitem[{Boone(2019)}]{Boone_2019}
Boone, K. 2019, The Astronomical Journal, 158, 257,
  \dodoi{10.3847/1538-3881/ab5182}

\bibitem[{Brout {et~al.}(2019)Brout, Sako, Scolnic, Kessler, D'Andrea, Davis,
  Hinton, Kim, Lasker, Macaulay, Möller, Nichol, Smith, Sullivan, Wolf, Allam,
  Bassett, Brown, Castander, Childress, Foley, Galbany, Herner, Kasai, March,
  Morganson, Nugent, Pan, Thomas, Tucker, Wester, Abbott, Annis, Avila, Bertin,
  Brooks, Burke, Rosell, Kind, Carretero, Crocce, Cunha, da~Costa, Davis,
  Vicente, Desai, Diehl, Doel, Eifler, Flaugher, Fosalba, Frieman, Garc{\'{\i}
  }a-Bellido, Gaztanaga, Gerdes, Goldstein, Gruen, Gruendl, Gschwend,
  Gutierrez, Hartley, Hollowood, Honscheid, James, Kuehn, Kuropatkin, Lahav,
  Li, Lima, Marshall, Martini, Miquel, Nord, Plazas, Roodman, Rykoff, Sanchez,
  Scarpine, Schindler, Schubnell, Serrano, Sevilla-Noarbe, Soares-Santos,
  Sobreira, Suchyta, Swanson, Tarle, Thomas, Tucker, Walker, Yanny, \&
  and}]{Brout_2019}
Brout, D., Sako, M., Scolnic, D., {et~al.} 2019, The Astrophysical Journal,
  874, 106, \dodoi{10.3847/1538-4357/ab06c1}

\bibitem[{Brout {et~al.}(2022)Brout, Scolnic, Popovic, Riess, Carr, Zuntz,
  Kessler, Davis, Hinton, Jones, Kenworthy, Peterson, Said, Taylor, Ali,
  Armstrong, Charvu, Dwomoh, Meldorf, Palmese, Qu, Rose, Sanchez, Stubbs,
  Vincenzi, Wood, Brown, Chen, Chambers, Coulter, Dai, Dimitriadis, Filippenko,
  Foley, Jha, Kelsey, Kirshner, Möller, Muir, Nadathur, Pan, Rest,
  Rojas-Bravo, Sako, Siebert, Smith, Stahl, \& Wiseman}]{pantheonplus}
Brout, D., Scolnic, D., Popovic, B., {et~al.} 2022, The Astrophysical Journal,
  938, 110, \dodoi{10.3847/1538-4357/ac8e04}

\bibitem[{Brunner {et~al.}(1997)Brunner, Connolly, Szalay, \&
  Bershady}]{Brunner_1997}
Brunner, R.~J., Connolly, A.~J., Szalay, A.~S., \& Bershady, M.~A. 1997, The
  Astrophysical Journal, 482, L21, \dodoi{10.1086/310674}

\bibitem[{Buchs {et~al.}(2019)Buchs, Davis, Gruen, DeRose, Alarcon, Bernstein,
  Sánchez, Myles, Roodman, Allen, Amon, Choi, Masters, Miquel, Troxel,
  Wechsler, Abbott, Annis, Avila, Bechtol, Bridle, Brooks, Buckley-Geer, Burke,
  Carnero Rosell, Carrasco Kind, Carretero, Castander, Cawthon, D’Andrea,
  da Costa, De Vicente, Desai, Diehl, Doel, Drlica-Wagner, Eifler, Evrard,
  Flaugher, Fosalba, Frieman, García-Bellido, Gaztanaga, Gruendl, Gschwend,
  Gutierrez, Hartley, Hollowood, Honscheid, James, Kuehn, Kuropatkin, Lima,
  Lin, Maia, March, Marshall, Melchior, Menanteau, Ogando, Plazas, Rykoff,
  Sanchez, Scarpine, Serrano, Sevilla-Noarbe, Smith, Soares-Santos, Sobreira,
  Suchyta, Swanson, Tarle, Thomas, Vikram, \& Collaboration)}]{SOM}
Buchs, R., Davis, C., Gruen, D., {et~al.} 2019, Monthly Notices of the Royal
  Astronomical Society, 489, 820, \dodoi{10.1093/mnras/stz2162}

\bibitem[{{Chen} {et~al.}(2022){Chen}, {Scolnic}, {Rozo}, {Rykoff}, {Popovic},
  {Kessler}, {Vincenzi}, {Davis}, {Armstrong}, {Brout}, {Galbany}, {Kelsey},
  {Lidman}, {M{\"o}ller}, {Rose}, {Sako}, {Sullivan}, {Taylor}, {Wiseman},
  {Asorey}, {Carr}, {Conselice}, {Kuehn}, {Lewis}, {Macaulay},
  {Rodriguez-Monroy}, {Tucker}, {Abbott}, {Aguena}, {Allam},
  {Andrade-Oliveira}, {Annis}, {Bacon}, {Bertin}, {Bocquet}, {Brooks}, {Burke},
  {Carnero Rosell}, {Carrasco Kind}, {Carretero}, {Cawthon}, {Costanzi}, {da
  Costa}, {Pereira}, {Desai}, {Diehl}, {Doel}, {Everett}, {Ferrero},
  {Flaugher}, {Friedel}, {Frieman}, {Garc{\'\i}a-Bellido}, {Gatti},
  {Gaztanaga}, {Gruen}, {Hinton}, {Hollowood}, {Honscheid}, {James}, {Lahav},
  {Lima}, {March}, {Menanteau}, {Miquel}, {Morgan}, {Palmese},
  {Paz-Chinch{\'o}n}, {Pieres}, {Plazas Malag{\'o}n}, {Prat}, {Romer},
  {Roodman}, {Sanchez}, {Schubnell}, {Serrano}, {Sevilla-Noarbe}, {Smith},
  {Soares-Santos}, {Suchyta}, {Tarle}, {Thomas}, {To}, {Tucker}, \&
  {Varga}}]{chen2022}
{Chen}, R., {Scolnic}, D., {Rozo}, E., {et~al.} 2022, \apj, 938, 62,
  \dodoi{10.3847/1538-4357/ac8b82}

\bibitem[{Dai {et~al.}(2018)Dai, Kuhlmann, Wang, \& Kovacs}]{Dai_2018}
Dai, M., Kuhlmann, S., Wang, Y., \& Kovacs, E. 2018, Monthly Notices of the
  Royal Astronomical Society, 477, 4142, \dodoi{10.1093/mnras/sty965}

\bibitem[{de~Oliveira {et~al.}(2022)de~Oliveira, Vargas~dos Santos, \&
  Reis}]{deOliveira:2022tvw}
de~Oliveira, F. M.~F., Vargas~dos Santos, M., \& Reis, R. R.~R. 2022, Mon. Not.
  Roy. Astron. Soc., 518, 2385, \dodoi{10.1093/mnras/stac3202}

\bibitem[{DeGroot \& Fienberg(1983)}]{degroot}
DeGroot, M.~H., \& Fienberg, S.~E. 1983, Journal of the Royal Statistical
  Society. Series D (The Statistician), 32, 12.
\newblock \url{http://www.jstor.org/stable/2987588}

\bibitem[{{D\'{}Isanto, A.} \& {Polsterer, K. L.}(2018)}]{disanto}
{D\'{}Isanto, A.}, \& {Polsterer, K. L.} 2018, A\&A, 609, A111,
  \dodoi{10.1051/0004-6361/201731326}

\bibitem[{{Frieman} {et~al.}(2008){Frieman}, {Bassett}, {Becker}, {Choi},
  {Cinabro}, {DeJongh}, {Depoy}, {Dilday}, {Doi}, {Garnavich}, {Hogan},
  {Holtzman}, {Im}, {Jha}, {Kessler}, {Konishi}, {Lampeitl}, {Marriner},
  {Marshall}, {McGinnis}, {Miknaitis}, {Nichol}, {Prieto}, {Riess}, {Richmond},
  {Romani}, {Sako}, {Schneider}, {Smith}, {Takanashi}, {Tokita}, {van der
  Heyden}, {Yasuda}, {Zheng}, {Adelman-McCarthy}, {Annis}, {Assef},
  {Barentine}, {Bender}, {Blandford}, {Boroski}, {Bremer}, {Brewington},
  {Collins}, {Crotts}, {Dembicky}, {Eastman}, {Edge}, {Edmondson}, {Elson},
  {Eyler}, {Filippenko}, {Foley}, {Frank}, {Goobar}, {Gueth}, {Gunn},
  {Harvanek}, {Hopp}, {Ihara}, {Ivezi{\'c}}, {Kahn}, {Kaplan}, {Kent},
  {Ketzeback}, {Kleinman}, {Kollatschny}, {Kron}, {Krzesi{\'n}ski}, {Lamenti},
  {Leloudas}, {Lin}, {Long}, {Lucey}, {Lupton}, {Malanushenko}, {Malanushenko},
  {McMillan}, {Mendez}, {Morgan}, {Morokuma}, {Nitta}, {Ostman}, {Pan},
  {Rockosi}, {Romer}, {Ruiz-Lapuente}, {Saurage}, {Schlesinger}, {Snedden},
  {Sollerman}, {Stoughton}, {Stritzinger}, {Subba Rao}, {Tucker}, {Vaisanen},
  {Watson}, {Watters}, {Wheeler}, {Yanny}, \& {York}}]{sdss}
{Frieman}, J.~A., {Bassett}, B., {Becker}, A., {et~al.} 2008, \aj, 135, 338,
  \dodoi{10.1088/0004-6256/135/1/338}

\bibitem[{Guo {et~al.}(2017)Guo, Pleiss, Sun, \& Weinberger}]{guo2017}
Guo, C., Pleiss, G., Sun, Y., \& Weinberger, K.~Q. 2017, On Calibration of
  Modern Neural Networks.
\newblock \doarXiv{1706.04599}

\bibitem[{{Guy} {et~al.}(2007){Guy}, {Astier}, {Baumont}, {Hardin}, {Pain},
  {Regnault}, {Basa}, {Carlberg}, {Conley}, {Fabbro}, {Fouchez}, {Hook},
  {Howell}, {Perrett}, {Pritchet}, {Rich}, {Sullivan}, {Antilogus}, {Aubourg},
  {Bazin}, {Bronder}, {Filiol}, {Palanque-Delabrouille}, {Ripoche}, \&
  {Ruhlmann-Kleider}}]{salt2}
{Guy}, J., {Astier}, P., {Baumont}, S., {et~al.} 2007, \aap, 466, 11,
  \dodoi{10.1051/0004-6361:20066930}

\bibitem[{He {et~al.}(2016)He, Zhang, Ren, \& Sun}]{he2016deep}
He, K., Zhang, X., Ren, S., \& Sun, J. 2016, in Proceedings of the IEEE
  conference on computer vision and pattern recognition, 770--778

\bibitem[{Hinton {et~al.}(2012)Hinton, Srivastava, Krizhevsky, Sutskever, \&
  Salakhutdinov}]{dropout}
Hinton, G.~E., Srivastava, N., Krizhevsky, A., Sutskever, I., \& Salakhutdinov,
  R.~R. 2012, Improving neural networks by preventing co-adaptation of feature
  detectors,  arXiv, \dodoi{10.48550/ARXIV.1207.0580}

\bibitem[{{Hounsell} {et~al.}(2018){Hounsell}, {Scolnic}, {Foley}, {Kessler},
  {Miranda}, {Avelino}, {Bohlin}, {Filippenko}, {Frieman}, {Jha}, {Kelly},
  {Kirshner}, {Mandel}, {Rest}, {Riess}, {Rodney}, \&
  {Strolger}}]{hounsell2018}
{Hounsell}, R., {Scolnic}, D., {Foley}, R.~J., {et~al.} 2018, \apj, 867, 23,
  \dodoi{10.3847/1538-4357/aac08b}

\bibitem[{{Ivezi{\'c}} {et~al.}(2019){Ivezi{\'c}}, {Kahn}, {Tyson}, {Abel},
  {Acosta}, {Allsman}, {Alonso}, {AlSayyad}, {Anderson}, {Andrew}, {Angel},
  {Angeli}, {Ansari}, {Antilogus}, {Araujo}, {Armstrong}, {Arndt}, {Astier},
  {Aubourg}, {Auza}, {Axelrod}, {Bard}, {Barr}, {Barrau}, {Bartlett}, {Bauer},
  {Bauman}, {Baumont}, {Bechtol}, {Bechtol}, {Becker}, {Becla}, {Beldica},
  {Bellavia}, {Bianco}, {Biswas}, {Blanc}, {Blazek}, {Blandford}, {Bloom},
  {Bogart}, {Bond}, {Booth}, {Borgland}, {Borne}, {Bosch}, {Boutigny},
  {Brackett}, {Bradshaw}, {Brandt}, {Brown}, {Bullock}, {Burchat}, {Burke},
  {Cagnoli}, {Calabrese}, {Callahan}, {Callen}, {Carlin}, {Carlson},
  {Chandrasekharan}, {Charles-Emerson}, {Chesley}, {Cheu}, {Chiang}, {Chiang},
  {Chirino}, {Chow}, {Ciardi}, {Claver}, {Cohen-Tanugi}, {Cockrum}, {Coles},
  {Connolly}, {Cook}, {Cooray}, {Covey}, {Cribbs}, {Cui}, {Cutri}, {Daly},
  {Daniel}, {Daruich}, {Daubard}, {Daues}, {Dawson}, {Delgado}, {Dellapenna},
  {de Peyster}, {de Val-Borro}, {Digel}, {Doherty}, {Dubois},
  {Dubois-Felsmann}, {Durech}, {Economou}, {Eifler}, {Eracleous}, {Emmons},
  {Fausti Neto}, {Ferguson}, {Figueroa}, {Fisher-Levine}, {Focke}, {Foss},
  {Frank}, {Freemon}, {Gangler}, {Gawiser}, {Geary}, {Gee}, {Geha}, {Gessner},
  {Gibson}, {Gilmore}, {Glanzman}, {Glick}, {Goldina}, {Goldstein}, {Goodenow},
  {Graham}, {Gressler}, {Gris}, {Guy}, {Guyonnet}, {Haller}, {Harris},
  {Hascall}, {Haupt}, {Hernandez}, {Herrmann}, {Hileman}, {Hoblitt}, {Hodgson},
  {Hogan}, {Howard}, {Huang}, {Huffer}, {Ingraham}, {Innes}, {Jacoby}, {Jain},
  {Jammes}, {Jee}, {Jenness}, {Jernigan}, {Jevremovi{\'c}}, {Johns}, {Johnson},
  {Johnson}, {Jones}, {Juramy-Gilles}, {Juri{\'c}}, {Kalirai}, {Kallivayalil},
  {Kalmbach}, {Kantor}, {Karst}, {Kasliwal}, {Kelly}, {Kessler}, {Kinnison},
  {Kirkby}, {Knox}, {Kotov}, {Krabbendam}, {Krughoff}, {Kub{\'a}nek},
  {Kuczewski}, {Kulkarni}, {Ku}, {Kurita}, {Lage}, {Lambert}, {Lange},
  {Langton}, {Le Guillou}, {Levine}, {Liang}, {Lim}, {Lintott}, {Long},
  {Lopez}, {Lotz}, {Lupton}, {Lust}, {MacArthur}, {Mahabal}, {Mandelbaum},
  {Markiewicz}, {Marsh}, {Marshall}, {Marshall}, {May}, {McKercher}, {McQueen},
  {Meyers}, {Migliore}, {Miller}, {Mills}, {Miraval}, {Moeyens}, {Moolekamp},
  {Monet}, {Moniez}, {Monkewitz}, {Montgomery}, {Morrison}, {Mueller},
  {Muller}, {Mu{\~n}oz Arancibia}, {Neill}, {Newbry}, {Nief}, {Nomerotski},
  {Nordby}, {O'Connor}, {Oliver}, {Olivier}, {Olsen}, {O'Mullane}, {Ortiz},
  {Osier}, {Owen}, {Pain}, {Palecek}, {Parejko}, {Parsons}, {Pease},
  {Peterson}, {Peterson}, {Petravick}, {Libby Petrick}, {Petry},
  {Pierfederici}, {Pietrowicz}, {Pike}, {Pinto}, {Plante}, {Plate}, {Plutchak},
  {Price}, {Prouza}, {Radeka}, {Rajagopal}, {Rasmussen}, {Regnault}, {Reil},
  {Reiss}, {Reuter}, {Ridgway}, {Riot}, {Ritz}, {Robinson}, {Roby}, {Roodman},
  {Rosing}, {Roucelle}, {Rumore}, {Russo}, {Saha}, {Sassolas}, {Schalk},
  {Schellart}, {Schindler}, {Schmidt}, {Schneider}, {Schneider}, {Schoening},
  {Schumacher}, {Schwamb}, {Sebag}, {Selvy}, {Sembroski}, {Seppala}, {Serio},
  {Serrano}, {Shaw}, {Shipsey}, {Sick}, {Silvestri}, {Slater}, {Smith},
  {Smith}, {Sobhani}, {Soldahl}, {Storrie-Lombardi}, {Stover}, {Strauss},
  {Street}, {Stubbs}, {Sullivan}, {Sweeney}, {Swinbank}, {Szalay}, {Takacs},
  {Tether}, {Thaler}, {Thayer}, {Thomas}, {Thornton}, {Thukral}, {Tice},
  {Trilling}, {Turri}, {Van Berg}, {Vanden Berk}, {Vetter}, {Virieux},
  {Vucina}, {Wahl}, {Walkowicz}, {Walsh}, {Walter}, {Wang}, {Wang}, {Warner},
  {Wiecha}, {Willman}, {Winters}, {Wittman}, {Wolff}, {Wood-Vasey}, {Wu},
  {Xin}, {Yoachim}, \& {Zhan}}]{ivezic_lsst}
{Ivezi{\'c}}, {\v{Z}}., {Kahn}, S.~M., {Tyson}, J.~A., {et~al.} 2019, \apj,
  873, 111, \dodoi{10.3847/1538-4357/ab042c}

\bibitem[{Kenworthy {et~al.}(2021)Kenworthy, Jones, Dai, Kessler, Scolnic,
  Brout, Siebert, Pierel, Dettman, Dimitriadis, Foley, Jha, Pan, Riess, Rodney,
  \& Rojas-Bravo}]{Kenworthy_2021}
Kenworthy, W.~D., Jones, D.~O., Dai, M., {et~al.} 2021, The Astrophysical
  Journal, 923, 265, \dodoi{10.3847/1538-4357/ac30d8}

\bibitem[{{Kessler} {et~al.}(2009){Kessler}, {Bernstein}, {Cinabro}, {Dilday},
  {Frieman}, {Jha}, {Kuhlmann}, {Miknaitis}, {Sako}, {Taylor}, \&
  {Vanderplas}}]{snana}
{Kessler}, R., {Bernstein}, J.~P., {Cinabro}, D., {et~al.} 2009, \pasp, 121,
  1028, \dodoi{10.1086/605984}

\bibitem[{Kessler {et~al.}(2010)Kessler, Cinabro, Bassett, Dilday, Frieman,
  Garnavich, Jha, Marriner, Nichol, Sako, Smith, Bernstein, Bizyaev, Goobar,
  Kuhlmann, Schneider, \& Stritzinger}]{Kessler_2010}
Kessler, R., Cinabro, D., Bassett, B., {et~al.} 2010, The Astrophysical
  Journal, 717, 40, \dodoi{10.1088/0004-637x/717/1/40}

\bibitem[{{Kessler} {et~al.}(2013){Kessler}, {Guy}, {Marriner}, {Betoule},
  {Brinkmann}, {Cinabro}, {El-Hage}, {Frieman}, {Jha}, {Mosher}, \&
  {Schneider}}]{sdss_simlib}
{Kessler}, R., {Guy}, J., {Marriner}, J., {et~al.} 2013, \apj, 764, 48,
  \dodoi{10.1088/0004-637X/764/1/48}

\bibitem[{{Kessler} {et~al.}(2019){Kessler}, {Narayan}, {Avelino}, {Bachelet},
  {Biswas}, {Brown}, {Chernoff}, {Connolly}, {Dai}, {Daniel}, {Di Stefano},
  {Drout}, {Galbany}, {Gonz{\'a}lez-Gait{\'a}n}, {Graham}, {Hlo{\v{z}}ek},
  {Ishida}, {Guillochon}, {Jha}, {Jones}, {Mand el}, {Muthukrishna}, {O'Grady},
  {Peters}, {Pierel}, {Ponder}, {Pr{\v{s}}a}, {Rodney}, \&
  {Villar}}]{plasticc_models}
{Kessler}, R., {Narayan}, G., {Avelino}, A., {et~al.} 2019, arXiv e-prints,
  arXiv:1903.11756.
\newblock \doarXiv{1903.11756}

\bibitem[{Kim \& Miquel(2007)}]{Kim_2007}
Kim, A.~G., \& Miquel, R. 2007, Astroparticle Physics, 28, 448,
  \dodoi{10.1016/j.astropartphys.2007.08.009}

\bibitem[{{Kingma} \& {Ba}(2014)}]{adam}
{Kingma}, D.~P., \& {Ba}, J. 2014, arXiv e-prints, arXiv:1412.6980.
\newblock \doarXiv{1412.6980}

\bibitem[{Krizhevsky {et~al.}(2017)Krizhevsky, Sutskever, \&
  Hinton}]{krizhevsky2017imagenet}
Krizhevsky, A., Sutskever, I., \& Hinton, G.~E. 2017, Communications of the
  ACM, 60, 84

\bibitem[{LeCun {et~al.}(1989)LeCun, Boser, Denker, Henderson, Howard, Hubbard,
  \& Jackel}]{lecun}
LeCun, Y., Boser, B., Denker, J., {et~al.} 1989, in Advances in Neural
  Information Processing Systems, ed. D.~Touretzky, Vol.~2 (Morgan-Kaufmann).
\newblock
  \url{https://proceedings.neurips.cc/paper/1989/file/53c3bce66e43be4f209556518c2fcb54-Paper.pdf}

\bibitem[{{LSST Science Collaboration} {et~al.}(2009){LSST Science
  Collaboration}, {Abell}, {Allison}, {Anderson}, {Andrew}, {Angel}, {Armus},
  {Arnett}, {Asztalos}, {Axelrod}, {Bailey}, {Ballantyne}, {Bankert},
  {Barkhouse}, {Barr}, {Barrientos}, {Barth}, {Bartlett}, {Becker}, {Becla},
  {Beers}, {Bernstein}, {Biswas}, {Blanton}, {Bloom}, {Bochanski}, {Boeshaar},
  {Borne}, {Bradac}, {Brandt}, {Bridge}, {Brown}, {Brunner}, {Bullock},
  {Burgasser}, {Burge}, {Burke}, {Cargile}, {Chandrasekharan}, {Chartas},
  {Chesley}, {Chu}, {Cinabro}, {Claire}, {Claver}, {Clowe}, {Connolly}, {Cook},
  {Cooke}, {Cooray}, {Covey}, {Culliton}, {de Jong}, {de Vries}, {Debattista},
  {Delgado}, {Dell'Antonio}, {Dhital}, {Di Stefano}, {Dickinson}, {Dilday},
  {Djorgovski}, {Dobler}, {Donalek}, {Dubois-Felsmann}, {Durech},
  {Eliasdottir}, {Eracleous}, {Eyer}, {Falco}, {Fan}, {Fassnacht}, {Ferguson},
  {Fernandez}, {Fields}, {Finkbeiner}, {Figueroa}, {Fox}, {Francke}, {Frank},
  {Frieman}, {Fromenteau}, {Furqan}, {Galaz}, {Gal-Yam}, {Garnavich},
  {Gawiser}, {Geary}, {Gee}, {Gibson}, {Gilmore}, {Grace}, {Green}, {Gressler},
  {Grillmair}, {Habib}, {Haggerty}, {Hamuy}, {Harris}, {Hawley}, {Heavens},
  {Hebb}, {Henry}, {Hileman}, {Hilton}, {Hoadley}, {Holberg}, {Holman},
  {Howell}, {Infante}, {Ivezic}, {Jacoby}, {Jain}, {R}, {Jedicke}, {Jee},
  {Garrett Jernigan}, {Jha}, {Johnston}, {Jones}, {Juric}, {Kaasalainen},
  {Styliani}, {Kafka}, {Kahn}, {Kaib}, {Kalirai}, {Kantor}, {Kasliwal},
  {Keeton}, {Kessler}, {Knezevic}, {Kowalski}, {Krabbendam}, {Krughoff},
  {Kulkarni}, {Kuhlman}, {Lacy}, {Lepine}, {Liang}, {Lien}, {Lira}, {Long},
  {Lorenz}, {Lotz}, {Lupton}, {Lutz}, {Macri}, {Mahabal}, {Mandelbaum},
  {Marshall}, {May}, {McGehee}, {Meadows}, {Meert}, {Milani}, {Miller},
  {Miller}, {Mills}, {Minniti}, {Monet}, {Mukadam}, {Nakar}, {Neill}, {Newman},
  {Nikolaev}, {Nordby}, {O'Connor}, {Oguri}, {Oliver}, {Olivier}, {Olsen},
  {Olsen}, {Olszewski}, {Oluseyi}, {Padilla}, {Parker}, {Pepper}, {Peterson},
  {Petry}, {Pinto}, {Pizagno}, {Popescu}, {Prsa}, {Radcka}, {Raddick},
  {Rasmussen}, {Rau}, {Rho}, {Rhoads}, {Richards}, {Ridgway}, {Robertson},
  {Roskar}, {Saha}, {Sarajedini}, {Scannapieco}, {Schalk}, {Schindler},
  {Schmidt}, {Schmidt}, {Schneider}, {Schumacher}, {Scranton}, {Sebag},
  {Seppala}, {Shemmer}, {Simon}, {Sivertz}, {Smith}, {Allyn Smith}, {Smith},
  {Spitz}, {Stanford}, {Stassun}, {Strader}, {Strauss}, {Stubbs}, {Sweeney},
  {Szalay}, {Szkody}, {Takada}, {Thorman}, {Trilling}, {Trimble}, {Tyson}, {Van
  Berg}, {Vanden Berk}, {VanderPlas}, {Verde}, {Vrsnak}, {Walkowicz},
  {Wandelt}, {Wang}, {Wang}, {Warner}, {Wechsler}, {West}, {Wiecha},
  {Williams}, {Willman}, {Wittman}, {Wolff}, {Wood-Vasey}, {Wozniak}, {Young},
  {Zentner}, \& {Zhan}}]{lsst_book}
{LSST Science Collaboration}, {Abell}, P.~A., {Allison}, J., {et~al.} 2009,
  arXiv e-prints, arXiv:0912.0201.
\newblock \doarXiv{0912.0201}

\bibitem[{Mitra {et~al.}(2022)Mitra, Kessler, More, Hlozek, \&
  Collaboration}]{mitra2022}
Mitra, A., Kessler, R., More, S., Hlozek, R., \& Collaboration, T. L. D. E.~S.
  2022, Using Host Galaxy Photometric Redshifts to Improve Cosmological
  Constraints with Type Ia Supernova in the LSST Era,  arXiv,
  \dodoi{10.48550/ARXIV.2210.07560}

\bibitem[{{M{\"o}ller} \& {de Boissi{\`e}re}(2020)}]{supernnova}
{M{\"o}ller}, A., \& {de Boissi{\`e}re}, T. 2020, \mnras, 491, 4277,
  \dodoi{10.1093/mnras/stz3312}

\bibitem[{Naeini {et~al.}(2015)Naeini, Cooper, \& Hauskrecht}]{naeini}
Naeini, M.~P., Cooper, G.~F., \& Hauskrecht, M. 2015, in Proceedings of the
  Twenty-Ninth AAAI Conference on Artificial Intelligence, AAAI'15 (AAAI
  Press), 2901–2907

\bibitem[{Niculescu-Mizil \& Caruana(2005)}]{niculescu}
Niculescu-Mizil, A., \& Caruana, R. 2005, in Proceedings of the 22nd
  International Conference on Machine Learning, ICML '05 (New York, NY, USA:
  Association for Computing Machinery), 625–632,
  \dodoi{10.1145/1102351.1102430}

\bibitem[{Palanque-Delabrouille {et~al.}(2010)Palanque-Delabrouille,
  Ruhlmann-Kleider, Pascal, Rich, Guy, Bazin, Astier, Balland, Basa, Carlberg,
  Conley, Fouchez, Hardin, Hook, Howell, Pain, Perrett, Pritchet, Regnault, \&
  Sullivan}]{Palanque_Delabrouille_2010}
Palanque-Delabrouille, N., Ruhlmann-Kleider, V., Pascal, S., {et~al.} 2010,
  Astronomy and Astrophysics, 514, A63, \dodoi{10.1051/0004-6361/200913283}

\bibitem[{Pasquet {et~al.}(2018)Pasquet, Bertin, Treyer, Arnouts, \&
  Fouchez}]{Pasquet_2018}
Pasquet, J., Bertin, E., Treyer, M., Arnouts, S., \& Fouchez, D. 2018,
  Astronomy {\&} Astrophysics, 621, A26, \dodoi{10.1051/0004-6361/201833617}

\bibitem[{{Perlmutter} {et~al.}(1999){Perlmutter}, {Aldering}, {Goldhaber},
  {Knop}, {Nugent}, {Castro}, {Deustua}, {Fabbro}, {Goobar}, {Groom}, {Hook},
  {Kim}, {Kim}, {Lee}, {Nunes}, {Pain}, {Pennypacker}, {Quimby}, {Lidman},
  {Ellis}, {Irwin}, {McMahon}, {Ruiz-Lapuente}, {Walton}, {Schaefer}, {Boyle},
  {Filippenko}, {Matheson}, {Fruchter}, {Panagia}, {Newberg}, {Couch}, \&
  {Project}}]{perlmutter}
{Perlmutter}, S., {Aldering}, G., {Goldhaber}, G., {et~al.} 1999, \apj, 517,
  565, \dodoi{10.1086/307221}

\bibitem[{{Pierel} {et~al.}(2018){Pierel}, {Rodney}, {Avelino}, {Bianco},
  {Filippenko}, {Foley}, {Friedman}, {Hicken}, {Hounsell}, {Jha}, {Kessler},
  {Kirshner}, {Mandel}, {Narayan}, {Scolnic}, \& {Strolger}}]{pierel2018}
{Pierel}, J.~D.~R., {Rodney}, S., {Avelino}, A., {et~al.} 2018, \pasp, 130,
  114504, \dodoi{10.1088/1538-3873/aadb7a}

\bibitem[{{Popovic} {et~al.}(2020){Popovic}, {Scolnic}, \&
  {Kessler}}]{popovic2020}
{Popovic}, B., {Scolnic}, D., \& {Kessler}, R. 2020, \apj, 890, 172,
  \dodoi{10.3847/1538-4357/ab6deb}

\bibitem[{{Qu} {et~al.}(2021){Qu}, {Sako}, {M{\"o}ller}, \& {Doux}}]{scone}
{Qu}, H., {Sako}, M., {M{\"o}ller}, A., \& {Doux}, C. 2021, \aj, 162, 67,
  \dodoi{10.3847/1538-3881/ac0824}

\bibitem[{Richard \& Lippmann(1991)}]{lippman}
Richard, M.~D., \& Lippmann, R.~P. 1991, Neural Computation, 3, 461,
  \dodoi{10.1162/neco.1991.3.4.461}

\bibitem[{{Riess} {et~al.}(1998){Riess}, {Filippenko}, {Challis},
  {Clocchiatti}, {Diercks}, {Garnavich}, {Gilliland}, {Hogan}, {Jha},
  {Kirshner}, {Leibundgut}, {Phillips}, {Reiss}, {Schmidt}, {Schommer},
  {Smith}, {Spyromilio}, {Stubbs}, {Suntzeff}, \& {Tonry}}]{riess}
{Riess}, A.~G., {Filippenko}, A.~V., {Challis}, P., {et~al.} 1998, \aj, 116,
  1009, \dodoi{10.1086/300499}

\bibitem[{Rigault {et~al.}(2020)Rigault, Brinnel, Aldering, Antilogus, Aragon,
  Bailey, Baltay, Barbary, Bongard, Boone, Buton, Childress, Chotard, Copin,
  Dixon, Fagrelius, Feindt, Fouchez, Gangler, Hayden, Hillebrandt, Howell, Kim,
  Kowalski, Kuesters, Leget, Lombardo, Lin, Nordin, Pain, Pecontal, Pereira,
  Perlmutter, Rabinowitz, Runge, Rubin, Saunders, Smadja, Sofiatti, Suzuki,
  Taubenberger, Tao, \& Thomas}]{Rigault_2020}
Rigault, M., Brinnel, V., Aldering, G., {et~al.} 2020, Astronomy {\&}
  Astrophysics, 644, A176, \dodoi{10.1051/0004-6361/201730404}

\bibitem[{Russakovsky {et~al.}(2015)Russakovsky, Deng, Su, Krause, Satheesh,
  Ma, Huang, Karpathy, Khosla, Bernstein, Berg, \& Fei-Fei}]{imagenet}
Russakovsky, O., Deng, J., Su, H., {et~al.} 2015, International Journal of
  Computer Vision (IJCV), 115, 211, \dodoi{10.1007/s11263-015-0816-y}

\bibitem[{{Sako} {et~al.}(2018){Sako}, {Bassett}, {Becker}, {Brown},
  {Campbell}, {Wolf}, {Cinabro}, {D'Andrea}, {Dawson}, {DeJongh}, {Depoy},
  {Dilday}, {Doi}, {Filippenko}, {Fischer}, {Foley}, {Frieman}, {Galbany},
  {Garnavich}, {Goobar}, {Gupta}, {Hill}, {Hayden}, {Hlozek}, {Holtzman},
  {Hopp}, {Jha}, {Kessler}, {Kollatschny}, {Leloudas}, {Marriner}, {Marshall},
  {Miquel}, {Morokuma}, {Mosher}, {Nichol}, {Nordin}, {Olmstead}, {{\"O}stman},
  {Prieto}, {Richmond}, {Romani}, {Sollerman}, {Stritzinger}, {Schneider},
  {Smith}, {Wheeler}, {Yasuda}, \& {Zheng}}]{sako2018}
{Sako}, M., {Bassett}, B., {Becker}, A.~C., {et~al.} 2018, \pasp, 130, 064002,
  \dodoi{10.1088/1538-3873/aab4e0}

\bibitem[{Simonyan \& Zisserman(2014)}]{Simonyan2014VeryDC}
Simonyan, K., \& Zisserman, A. 2014, CoRR, abs/1409.1556

\bibitem[{Sønderby {et~al.}(2020)Sønderby, Espeholt, Heek, Dehghani, Oliver,
  Salimans, Agrawal, Hickey, \& Kalchbrenner}]{metnet}
Sønderby, C.~K., Espeholt, L., Heek, J., {et~al.} 2020, MetNet: A Neural
  Weather Model for Precipitation Forecasting.
\newblock \doarXiv{2003.12140}

\bibitem[{{The PLAsTiCC team} {et~al.}(2018){The PLAsTiCC team}, {Allam},
  {Bahmanyar}, {Biswas}, {Dai}, {Galbany}, {Hlo{\v{z}}ek}, {Ishida}, {Jha},
  {Jones}, {Kessler}, {Lochner}, {Mahabal}, {Malz}, {Mandel},
  {Mart{\'\i}nez-Galarza}, {McEwen}, {Muthukrishna}, {Narayan}, {Peiris},
  {Peters}, {Ponder}, {Setzer}, {The LSST Dark Energy Science Collaboration},
  {LSST Transients}, \& {Variable Stars Science Collaboration}}]{plasticc_data}
{The PLAsTiCC team}, {Allam}, Tarek, J., {Bahmanyar}, A., {et~al.} 2018, arXiv
  e-prints, arXiv:1810.00001.
\newblock \doarXiv{1810.00001}

\bibitem[{van~den Oord {et~al.}(2016)van~den Oord, Kalchbrenner, \&
  Kavukcuoglu}]{pixelrnn}
van~den Oord, A., Kalchbrenner, N., \& Kavukcuoglu, K. 2016, Pixel Recurrent
  Neural Networks.
\newblock \doarXiv{1601.06759}

\bibitem[{{Vincenzi} {et~al.}(2023){Vincenzi}, {Sullivan}, {M{\"o}ller},
  {Armstrong}, {Bassett}, {Brout}, {Carollo}, {Carr}, {Davis}, {Frohmaier},
  {Galbany}, {Glazebrook}, {Graur}, {Kelsey}, {Kessler}, {Kovacs}, {Lewis},
  {Lidman}, {Malik}, {Nichol}, {Popovic}, {Sako}, {Scolnic}, {Smith}, {Taylor},
  {Tucker}, {Wiseman}, {Aguena}, {Allam}, {Annis}, {Asorey}, {Bacon}, {Bertin},
  {Brooks}, {Burke}, {Carnero Rosell}, {Carretero}, {Castander}, {Costanzi},
  {da Costa}, {Pereira}, {De Vicente}, {Desai}, {Diehl}, {Doel}, {Everett},
  {Ferrero}, {Flaugher}, {Fosalba}, {Frieman}, {Garc{\'\i}a-Bellido}, {Gerdes},
  {Gruen}, {Gutierrez}, {Hinton}, {Hollowood}, {Honscheid}, {James}, {Kuehn},
  {Kuropatkin}, {Lahav}, {Li}, {Lima}, {Maia}, {Marshall}, {Miquel}, {Morgan},
  {Ogando}, {Palmese}, {Paz-Chinch{\'o}n}, {Pieres}, {Plazas Malag{\'o}n},
  {Reil}, {Roodman}, {Sanchez}, {Schubnell}, {Serrano}, {Sevilla-Noarbe},
  {Suchyta}, {Tarle}, {To}, {Varga}, {Weller}, {Wilkinson}, \& {DES
  Collaboration}}]{vincenzi_contamination}
{Vincenzi}, M., {Sullivan}, M., {M{\"o}ller}, A., {et~al.} 2023, \mnras, 518,
  1106, \dodoi{10.1093/mnras/stac1404}

\bibitem[{{Wang} {et~al.}(2015){Wang}, {Gjergo}, \& {Kuhlmann}}]{wang2015}
{Wang}, Y., {Gjergo}, E., \& {Kuhlmann}, S. 2015, \mnras, 451, 1955,
  \dodoi{10.1093/mnras/stv1090}

\bibitem[{Zeiler \& Fergus(2014)}]{zeiler2014visualizing}
Zeiler, M.~D., \& Fergus, R. 2014, in European conference on computer vision,
  Springer, 818--833

\end{thebibliography}
\appendix
\section{Survey-Agnostic Model Performance}
\begin{figure}[h!]
    \centering
    \includegraphics[scale=0.47]{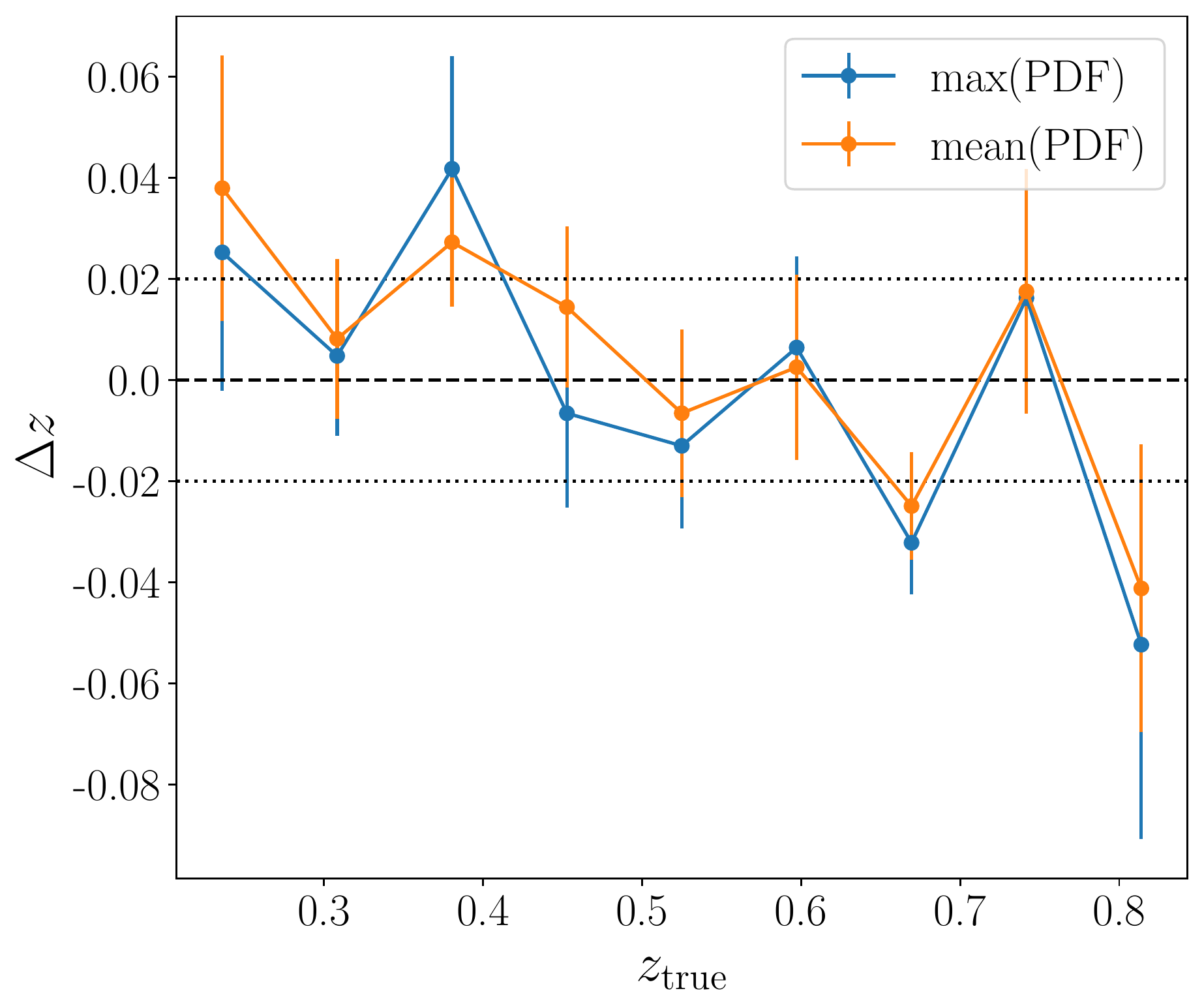}
    \caption{Mean binned residuals, $\Delta z \equiv \frac{z_{\mathrm{pred}} - z_{\mathrm{true}}}{1+z_{\mathrm{true}}}$, as a function of true redshift, $z_{\mathrm{true}}$, for the DES3YR SNe Ia sample produced by a model trained on the PLAsTiCC dataset. The max(PDF) and mean(PDF) methods of obtaining point estimates from \name\ PDFs are described in \S\ref{subsec:metrics}.}
    \label{fig:des-test}
\end{figure}

Due to the 2D Gaussian process regression in both wavelength and time dimensions that is performed in order to create the input images (details in \S\ref{subsec:lc-preprocess}), data from different surveys with different photometric bands can be processed into the same image format. This creates the potential for a survey-agnostic model that can be trained on data from a single survey and applied to data from others. To test this hypothesis, we evaluated our model trained on PLAsTiCC data, described in \S\ref{subsec:plasticc}, on 248 spectroscopically confirmed SNe Ia from the first 3 years of the Dark Energy Survey supernova program \citep[DES3YR,][]{Brout_2019}. We show these results in Figure~\ref{fig:des-test}.

The scatter and redshift-dependent bias are certainly more prominent in this result than the others presented in this work (Figures~\ref{fig:plasticc-resids} and~\ref{fig:sdss-delta-z}), but is nevertheless an impressive result considering that this 2D Gaussian process regression method uniquely makes cross-survey results possible. We also highlight that these observed SNe Ia cover a much larger redshift range than the SDSS observed sample, demonstrating the applicability of \name\ on high-redshift, deep sky surveys such as LSST. While training a fresh model with a full suite of Dark Energy Survey SNe Ia simulations would certainly produce improved results, we posit that improvements could be attained relatively inexpensively by "fine-tuning" an existing trained model using a small volume of SNe Ia simulated in the target survey. We leave these interesting extensions to future work.

\section{Redshift Prediction Outliers of the Real SDSS Dataset}
\begin{figure}
    \centering
    \includegraphics[scale=0.47]{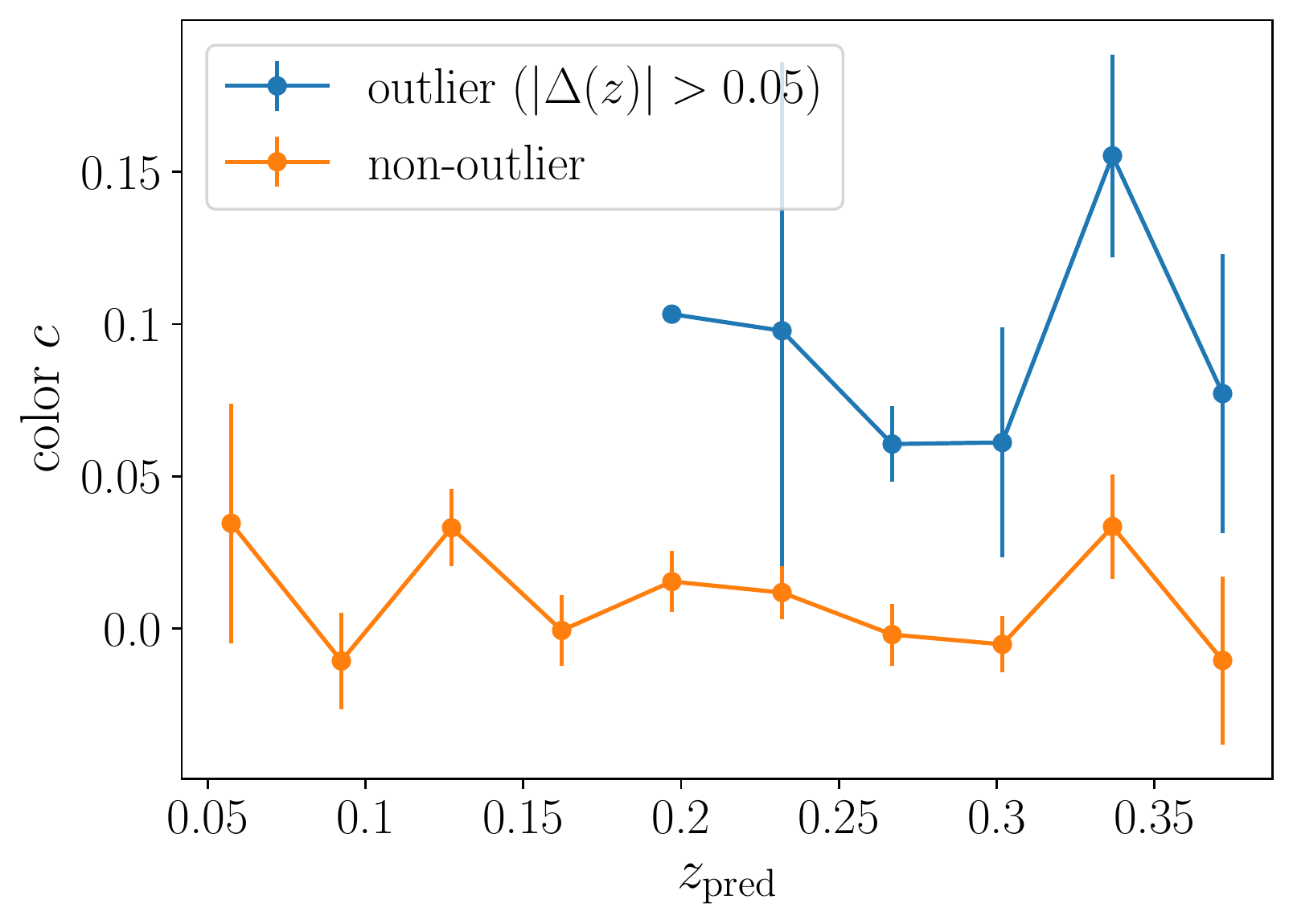}
    \caption{Mean binned SALT2 color, $c$, as a function of predicted redshift, $z_{\mathrm{pred}}$, for the \name\ prediction outliers and non-outliers in the SDSS observed photometric SN Ia sample.}
    \label{fig:outliers}
\end{figure}

We investigate the 29 \name\ prediction outliers, defined as objects with $\Delta z > 0.05$, in the observed SDSS dataset. We find that the population of outliers is noticeably redder than their non-outlier counterparts at the same predicted redshift (Figure~\ref{fig:outliers}), which could explain the prediction failure.

We also attempted to use the predicted PDF shapes to remove outliers, assuming that poor predictions may be correlated with wider PDFs, representing the model's lack of confidence in less accurate predictions. Although we found a slight correlation, it is not strong enough for this metric to be useful: 48\% of outliers have "wide" PDFs (defined as $\sigma > 0.025$ empirically) as well as 18\% of non-outliers.

\end{document}